\documentclass[review,final]{elsarticle}

\journal{Journal of Sound and Vibration}

\usepackage[utf8]{inputenc}
\usepackage[svgnames]{xcolor}
\usepackage[english]{babel}
\usepackage[mathscr]{eucal}
\usepackage{graphicx}
\usepackage{amsmath}
\usepackage{amsthm}
\usepackage{amssymb}
\usepackage{amsfonts}
\usepackage[colorlinks,unicode,allcolors=DarkCyan,pdftitle={Moving oscillator}]{hyperref}

\usepackage{listings}
\usepackage{lastpage}
\usepackage{afterpage}
\usepackage{setspace}
\usepackage{mathtools}

\def\d{\mathrm d}

\def\const{\operatorname {const}}
\def\sign{\operatorname {sign}}

\def\sign{\operatorname{sign}}
\def\const{\operatorname{const}}

\def\pd#1#2{\dfrac{\partial#1}{\partial#2}}

\def\tauu{\tau}

\def\I{\mathrm i}

\definecolor{brown}{RGB}{163,40,40}

\newtheorem{proposition}{Proposition}
\theoremstyle{remark}
\newtheorem{remark}{Remark}

\def\upsilon{v}
\def\nu{}

\renewcommand{\=}{\stackrel{\mbox{\scriptsize def}}{=}}

\renewcommand{\d}{\text{d}}

\renewcommand{\U}{\mathscr{U}}

\makeatletter

\@addtoreset{equation}{section}
\makeatother

\begin{document}
\begin{frontmatter}

\selectlanguage{english}
\title{Non-stationary oscillation of a string on the Winkler foundation subjected to
a discrete mass-spring system non-uniformly moving at a sub-critical speed}

\author[ipme]{Serge N.~Gavrilov\corref{mycorrespondingauthor}} 
\ead{serge@pdmi.ras.ru}
\cortext[mycorrespondingauthor]{Corresponding author}
\author[ipme]{Ekaterina V.~Shishkina}
\ead{shishkina\_k@mail.ru}
\author[spbstu]{Ilya O.~Poroshin}
\ead{poroshin\_io@mail.ru}
\address[ipme]{Institute for Problems in Mechanical Engineering RAS, V.O., Bolshoy
pr. 61, St.~Petersburg, 199178, Russia}
\address[spbstu]{Peter the Great St.~Petersburg Polytechnic University (SPbPU),
Polytechnicheskaya str.~29, St.Petersburg, 195251, Russia}



\selectlanguage{english}

\begin{abstract}
We consider non-stationary free and forced transverse oscillation of an
infinite taut string
on the Winkler foundation subjected to a discrete mass-spring system non-uniformly
moving at a given sub-critical speed. The speed of the mass-spring system is assumed
to be a slowly time-varying function. 
To describe a non-vanishing free oscillation
we use an analytic approach based on the method of stationary phase and
the method of multiple scales. 
The moving oscillator is characterized by a partial frequency, which can be
greater or less than the cut-off frequency.
Accordingly, a sub-critical uniformly
accelerated motion generally has two stages. At the first stage there exists
a trapped mode, and, therefore, a part of the wave energy 
is localized near the moving oscillator and does not propagate away.
For this stage we obtain the analytic solution in a simple
form describing non-vanishing free oscillation and verify it numerically. For the second stage there is no trapped
mode, and all the wave energy propagate away. This stage is investigated numerically, and some unexpected results are
obtained. 
Additionally, we consider the case of the oscillator with a destabilizing spring. The dynamics of the system in the latter case is quite different from the case of commonly
used stabilizing spring, since the system loses the stability during an accelerated motion.
We also take into consideration the forced
oscillation caused by an external load being a superposition of harmonics with
time-varying parameters (the amplitude and the frequency). 
\end{abstract}


\end{frontmatter}
\section{Introduction}
\begin{figure}[htp]
\centering{\includegraphics[width=0.8\textwidth]{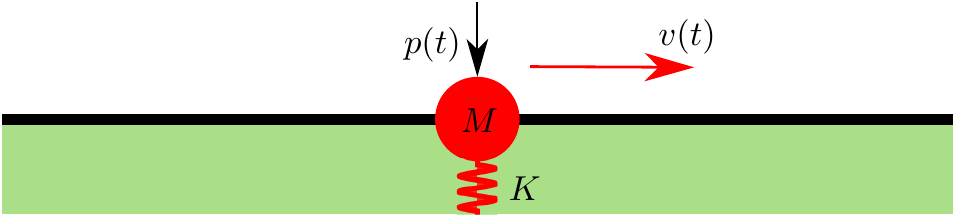}
{%
}
}
\caption{The schematic of the system}
\label{winkler-spring.eps}
\end{figure}

In the paper we deal with a moving load problem
\cite{fryba1972vibration,Vesnitskiy2001eng}. 
We consider transverse oscillation of an infinite taut string on the Winkler 
foundation. The string is equipped with a moving discrete mass-spring
oscillator non-uniformly moving at a given sub-critical speed 
(see Fig.~\ref{winkler-spring.eps} for the schematic of the system).

Most of the results concerning non-uniformly moving loads applied to {a
string} are obtained for the
case of an inertialess point loads
\cite{Flaherty:1968,Stronge:1971,kaplunov1986vibrations,Non-Stationary}.
A famous effect related to a point inertial load was demonstrated by Stokes
\cite{stokes1849discussion} who considered a uniform motion of a point mass along
a finite inertialess string and obtained a paradoxical result 
concerning
discontinuity in the particle trajectory close to the end support
(see also \cite{fryba1972vibration,smith1964motions,dyniewicz2009paradox}).
The Stokes paradox was resolved in recent series of studies 
\cite{gavrilov2016revisitation,Ferretti2019,FerrettiJSV2019}
where the dynamic problem in the framework of geometrically non-linear problem
statement was investigated following to ideas of 
\cite{rayleigh1902xxxiv,Nicolai1912-eng,nicolai1925xix,vesnitski1983laws,Gavrilov-2006-trans}.
Motion of a small point mass along a semi-infinite string without elastic
foundation was considered in \cite{Rodeman:1976}. 
Motion of a linear oscillator along a finite string or beam is considered in 
\cite{Yang2000,Pesterev1997,Pesterev1997a} mostly from the numerical point of
view. Analytical solutions for infinite, semi-infinite and finite strings
without elastic foundation with uniformly moving masses are obtained in 
\cite{Gao2015}.
The nonlinear problem concerning a moving mass along a finite suspended cable was
considered in \cite{Wang2010}.
Instability of transversal vibration of a mass moving along a string on
the elastic foundation with periodic or stochastic near-periodic 
stiffness was considered in 
\cite{Metrikine1996,vesnitskii1996instability}.

In \cite{gavrilov2002etm} it was suggested a new approach based on
the method of stationary phase and the method of multiple scales successively
applied to the problem concerning
a non-uniform sub-critical motion of a point mass along an infinite string on the Winkler
foundation. The method of multiple scales was applied to describe the evolution of
the amplitude of a trapped mode of oscillation, due to which a non-stationary 
free oscillation in such a system does not vanish with time. To the best of
our knowledge, such an approach was never used before to investigate the dynamics
of systems with moving loads.
The extensive bibliography on the phenomenon of trapped modes and localization
of linear waves can be found in
\cite{Ind-book-R2E,arXiv:1805.07382,kaplunov2008example,Mishuris2020}. This method, which is not
a pure asymptotic one, has a deep
asymptotic motivation (see Sect.~\ref{sect-method} for the discussion).
It is a general approach, which allows us to investigate non-stationary free
localized oscillation in infinite systems with time-varying parameters in the
case if the corresponding system with constant parameters possesses a single
trapped mode. The method was applied to several model problems
concerning an infinite string on elastic foundation with a discrete inclusion,
namely a string with
time-varying tension \cite{arXiv:1805.07382}, string with time-varying mass
\cite{indeitsev2016evolution}, string with oscillator of time-varying
stiffness \cite{gavrilov-da70}. Also, the method was successfully applied to
the problem concerning a beam with discrete inclusion of a time-varying stiffness
\cite{shishkina2018non},
and a resonant solution of a model problem concerning a string 
\cite{Shishkina2020}
was obtained. In all cases we compare
our analytic results with numerics and demonstrate an excellent agreement.
This allows us to develop some mathematical tricks, which simplify the problem
solution.

In this paper we deal with a moving discrete sub-system, namely a moving
mass-spring system (a linear oscillator). 
The moving oscillator is characterized by a partial frequency, which can be
greater or less than the cut-off frequency for the infinite waveguide (i.e., the
string on elastic foundation). 
Accordingly, a sub-critical uniformly
accelerated motion generally has two stages. At the first stage there exists
the trapped mode, and, therefore, a part of the wave energy 
is localized near the moving oscillator and does not propagate away.
For this stage we obtain the analytic solution in a simple
form and verify it numerically. For the second stage there is no trapped
mode, and all the wave energy propagate away. This stage is investigated numerically, and some unexpected results are
obtained (the growth of the force between the load and the string, see 
Sect~\ref{sec-free-gg} for details). We also provide some ideas how the second stage
can be described analytically (though this remains to be a subject of a future
work, see Sect.~\ref{sec-conclusion}).  To the best of our knowledge, these two stages of
an accelerated motion are not identified and described
in the modern literature. Moreover, the corresponding two qualitatively different
cases (whether or not there exist trapped modes) in the problem concerning a uniform sub-critical motion
are also not discussed (see, e.g., recent study \cite{Roy2018}).

We underline that complicated models involving an infinite waveguide and several moving 
discrete inclusions constructed by spring, mass, and damper elements 
\cite{Roy2018}, which are used in engineering to investigate, for
example, pantograph-catenary dynamics
\cite{Metrikine2006,VoVan2017,JimenezOctavio2015,PilJung2012,Gil2020},
under certain conditions definitely 
can possess trapped modes in the absence of viscosity in dampers, and,
therefore, a non-vanishing free oscillation can
be observed there. Note that adding of damper elements with a small viscosity
does not change the situation significantly, since
in the latter
case free localized oscillation vanishes very slowly \cite{kaplunov1986torsional}.
A damper element is not taken into account in the current paper, though for the
time being we have some ideas for future how to consider it in the framework of
our approach. The obtained analytic solution can be used as a reference
solution in applications, and brings the understanding that a sub-critical
accelerated motion of a discrete system composed by several masses, springs,
and dampers 
can have qualitatively different stages.

The problem considered in \cite{gavrilov2002etm} is a limiting case of the current
problem. The second stage of an accelerated motion does not exist for a point mass
problem investigated in that previous paper. Note that in \cite{gavrilov2002etm}
 the solution initially was obtained in extremely
complicated form as a product of ten indefinite integrals, though the final
result has a very simple structure. In this paper we suggest a mathematical trick,
which allows us
to significantly simplify the calculations and to obtain the solution of more
complicated problems in an easier way. 
While considering the corresponding limiting case, we have discovered an error in
\cite{gavrilov2002etm}, though the behaviour of the erroneous solution is
very close to the correct solution (see Sect.~\ref{sect-comparison-old} for
details). Note that revisitation of \cite{gavrilov2002etm} has never been a
motivation of our current work, since we have been absolutely sure that the old
solution is correct being in an excellent agreement with numerical results.
In addition to the fact that now we deal with a mechanical system different from the one
considered in \cite{gavrilov2002etm},
in this paper
we introduce some other extensions to the problem formulation, namely:
\begin{itemize} 
\item 
We extend the class of admissible loads applied to the discrete subsystem, for
which the analytic solution can be obtained. In \cite{gavrilov2002etm} the
external load is the 
constant weight of the point mass.
Now we introduce a load, which can be represented 
as the superposition of a ``pulse'' load, which
acts during some time, and a number of harmonics with independent (from the
speed) time-varying parameters (the frequency and the amplitude).
Thus, the forced
oscillation, caused by such a harmonic load, co-existing with a free localized oscillation
is under investigation.
The obtained analytic solution is verified numerically.

\item 
We consider both the classical case of a stabilizing oscillator spring, and the case when
the stiffness of the spring in the oscillator is negative (i.e.,\ we deal with
a destabilizing spring).
Destabilizing springs can be realized by magnetic forces 
\cite{oyelade2017dynamics}.
They are used, e.g., 
in multimode control of cable vibration 
\cite{Chen2021,Chen2015}
and when constructing metamaterials 
\cite{grekova2018dd,chronopoulos2015enhancement,pasternak2014chains}.
The dynamics of the system in the latter case is quite different from the case of
commonly used stabilizing spring, since the system loses the stability during
an accelerated motion.
\end{itemize}

There are only few known analytic solutions concerning non-stationary
oscillation of infinite waveguides caused by non-uniformly moving inertial
loads. To the best of our knowledge, our previous paper \cite{gavrilov2002etm} is the only one
paper, where such a solution is obtained in a simple asymptotic form. 
Almost all results concerning moving loads in infinite dispersive waveguides 
are obtained by applying the method of integral transforms,
which generally cannot yield simple analytic results for systems with
time-varying parameters. We have suggested a general alternative
approach based on the method of multiple scales. On the other hand, as far as
we know, in the modern literature there are no analytic solutions for a
system with a non-uniformly moving
oscillator, which demonstrates more complicated and qualitatively different
dynamic behaviour.  
Thus, we suppose that our new results are an important contribution to the
understanding of dynamics for systems with moving loads. 

The paper is organized as follows.
In Sect.~\ref{sec-formulation} we present mathematical formulation for the
problem in the fixed (non-moving) co-ordinates (that are used in the
paper in the numerical treatment), 
as well as in the co-moving with the load co-ordinates (that are used in the
paper in the analytic treatment).
In Sect.~\ref{sect-method} we discuss our analytic approach. We briefly
discuss what a trapped mode is, and why it is important for our problem.
We introduce a class of the admissible loadings applied to the mass-spring
system.
Thus, in this
paper we deal with both free and forced oscillation, and, accordingly, we need to
introduce some modifications to our method. The solution is represented in the
form that we call the multi-frequency ansatz
(Sect.~\ref{sect-multi-frequency-ansatz}), which selects a number of harmonics
with time-varying frequencies considered to be important. The choice of these
frequencies is based on the results got by the method of stationary
phase 
\cite{fedoruk1977,temme2014} applied to the corresponding system with constant parameters.
Every term of the multi-frequency ansatz is represented in
the form of an asymptotic series, which we call the single-frequency ansatz 
(Sect.~\ref{sect-single-frequency-ansatz}). The aim of the analytic work in
this paper is to find the principal term of the multi-frequency ansatz. To do this
we use a modification of the method of multiple scales 
\cite{feschenko1967eng,nayfehperturbation}. In Sect.~\ref{sec-diff-op} we
present the multi-scale representation for the differential operators.
In Sect.~\ref{sec-asymptotic} we
obtain the analytic solution for our problem.
In Sect.~\ref{sec-c-trapped} we evaluate free localized oscillation, and this
is the most important part of the paper. In Sect.~\ref{sec-c-forced} we evaluate
the modes of forced oscillation. In Sect.~\ref{sec-c-constants} we calculate the
unknown constants in the expression describing the  free
oscillation. In Sect.~\ref{sect-c-iforce}
we obtain the analytic solution for the unknown internal force between the
string and the discrete mass. In Sect.~\ref{sect-numerics} we verify the
constructed analytic solution numerically. To do this we derive an integral
equation for the unknown internal force (Sect.~\ref{sec-i-eq}), 
and solve it numerically to compare the results (Sect.~\ref{sec-comparison})
for several qualitatively different cases.
 The cases of pure free oscillation for a stabilizing
and destabilizing discrete oscillator spring are considered in 
Sects.~\ref{sec-free-gg}, \ref{sec-free-l}, respectively. 
In Sect.~\ref{sect-comparison-old} we compare obtained results with results of
previous paper \cite{gavrilov2002etm} and demonstrate that the old solutions
is erroneous, though it has a behaviour, which is very close to the
correct solution.
In Sect.~\ref{sec-not-exist} we demonstrate numerically that a free
oscillation is negligible if the trapped mode does not exist in the system, as
generally expected.
In Sect.~\ref{sec-forced} we verify the analytic solution in the case of
co-existing free and forced oscillation. 
In Sect.~\ref{sec-final-remarks} we very briefly discuss some more qualitatively
different cases. In Conclusion (Sect.~\ref{sec-conclusion}) we discuss the
basic results of the paper and its possible generalizations.

The case of a uniform motion of the
mass-spring oscillator is not considered to be the subject of this paper. 
Thus, all necessary auxiliary results are presented in Appendix. The material
of \ref{App-A}--\ref{spectral-joined}
mostly involves some known results or their 
generalizations (see
\cite{Kruse1998,gavrilov2002etm,
arXiv:1805.07382,
gavrilov-da70,Ind-book-R2E,Glushkov2011a,gao2014exact}).
The final result 
(\ref{neodnor})
concerning the large-time asymptotics of
non-stationary free and forced oscillation in the system with constant
parameters is the generalization of the results obtained in 
\cite{kaplunov1986torsional}, though we use a bit different technique
for the asymptotic evaluation of integrals. Finally, in
\ref{app-f-s}--\ref{app-f-s1} we present the formulae for the
fundamental solutions in time domain, which we use to derive the integral
equation.

\section{Mathematical formulation}
\label{sec-formulation}
Introduce the following notation: $u(x,t)$ is the displacement of a point of the string at the
position $x$ and the time $t$, 
$M \geq 0$ is the mass in the discrete oscillator, $K$ is the spring
stiffness for the discrete oscillator.
We assume that $K$ can be positive (stabilizing), {negative
(destabilizing)}, or zero:
\begin{equation}
       K\lesseqqgtr0. 
\end{equation}
As we have already discussed in Introduction
destabilizing springs are used when constructing metamaterials.

For $t<0$ the string and the discrete oscillator are at the rest. 
At the instant $t=0$ an external non-zero transverse force $p(t)$ applied to
the discrete mass appears, and the oscillator starts to
move along the string according to the law  
\begin{equation}
\ell(t)=\ell(0)+\int_{0}^t v(T)\,\d T,
\end{equation}
where $v(t)$ is the speed of the oscillator (a given
function), and $\ell (0)$ is a given initial position for the oscillator.
Denote the acceleration of the oscillator as  
$a(t)= \dot{\upsilon}(t)$. We consider a sub-critical
regime of the oscillator motion and assume that for all $t$
\begin{equation}
\big|v(t)\big|<1.
\label{v-is-subcritical}
\end{equation}
Thus, the motion $\ell(t)$ of the oscillator along the string and the
transverse 
external force $p(t)$ are considered as prescribed quantities. The transverse
dynamics of the system is unknown.

Denote the displacement of the point
of the string subjected to the moving load as 
\begin{equation}
\mathscr{U}(t)=u(\ell(t),t). 
\end{equation}
The governing equations in the dimensionless form
\cite{gavrilov-da70}
are
\begin{gather}
\pd{{}^2u}{{x}^2}-
\pd{{}^2u}{{t}^2}-
 u =  -P({t}) \delta
\big({x}-{\ell}({t})\big), 
\label{ur_str_2-dim}
\\
	M \frac{d^2\mathscr{U}}{dt^2} + K
        {\mathscr{U}}
        = -P({t}) +p({t}) ,
\label{ur_osc_2}
\end{gather}
where
$P(t)$ is the unknown transverse force on the string from the oscillator,
$\delta$ is the Dirac delta-function.
Equations~\eqref{ur_str_2-dim},
\eqref{ur_osc_2}
can be rewritten in the co-moving with the discrete oscillator system of co-ordinates
$\xi = x-\ell(\tau)$, $\tau=t$:
\begin{gather}
\label{ur_str_3}
(1-\upsilon^2)u''
+a u'
+2\upsilon 
\dot u'-\ddot u-u =  -P(\tau) \delta(\xi),
\\
M \ddot{\mathscr{U}}
+ K \mathscr{U}
= -P(\tau) +p(\tau).
\label{ur_osc_2-dupe}
\end{gather}
Here and in what follows, we denote by prime and overdot the derivatives with respect to
$\xi$ and $\tau$, respectively.

According to Eqs.~\eqref{ur_str_2-dim}, \eqref{ur_osc_2}
the following Hugoniot conditions must be satisfied
under the moving load:
\begin{gather}
[u] 
=0,
\label{usl_Gu_1}
\\
[u'] 
=-\frac{P(\tau)}{1-\upsilon^2}.
\label{usl_Gu_2}
\end{gather}
Here, and in what follows, $[\mu]\equiv\mu(\xi+0)-\mu(\xi-0)$ for any
arbitrary quantity $\mu(\xi, \tau)$.

The initial conditions for Eq.~\eqref{ur_str_3} are zero.
They can be
formulated in the following form, which is conventional for distributions (or
generalized functions) \cite{Vladimirov1971}:
\begin{equation}
	u \big|_{\tau<0} \equiv 0.
\label{initial-cond}
\end{equation}	
	
%

%
%

\section{Method}
\label{sect-method}
The main assumption we use to construct the asymptotic solution is that 
the velocity $v$
is a slowly
varying piecewise monotone function of the slow time $T$:
\begin{gather}
T=\epsilon\tau,
\\
v=v(T), 
\end{gather}
and, therefore, acceleration
$a=O(\epsilon)$.
Here $\epsilon$ is a formal small parameter. 
We look for the asymptotic solution under the following conditions:
\begin{itemize}
\item 
$
\epsilon=o(1), 
$
\item
$\tau=O(\epsilon^{-1}),$
\item
$v(T)$ satisfies {restriction}
\eqref{v-is-subcritical}
for all $T$. Note that in what follows, we will formulate more strict restrictions,
see Eqs.~\eqref{v-Kgt0}, \eqref{v-Kless0}.
\end{itemize}
In the paper we restrict ourselves to the external loadings $p$, which 
can be represented as a superposition of a ``pulse'' loading, which
acts during some time, and a number of harmonics with slowly time-varying
frequencies and  amplitudes:
\begin{equation}
  p(\tau,T)=H(\tau)\left(\frac{\hat p(\tau)}2+\sum_{i=1}^Np^{(\Omega_i)}(T)\exp\left(-\I\int_0^\tau
  \Omega_i(T)\,\d T
  \right)\right)+\mathrm{c.c.},
\label{p-def}
\end{equation}
where notation $\mathrm{c.c.}$ denotes 
the complex conjugate terms for the whole right-hand side, $H(\cdot)$ is the
Heaviside function,
\begin{gather}
\Omega_i\geq0,
\\
\Omega_i\not\simeq\Omega_j \quad \forall t\quad
\text{if} \quad{i\neq j}\quad\text{for}\quad{i,j=\overline{1,N}}.
\label{non-resonant}
\end{gather}
We assume the pulse loading 
{$\hat p(\tau)$ to be a given real finite integrable function (or a finite generalized
function \cite{Lighthill1964,Vladimirov1971}) such that 
$\hat p\equiv0$ if $\tau<0$ or $\tau>\mathscr T$} for certain $\mathscr
T>0$. 
The amplitudes $p^{(\Omega_i)}(T)\ (i=\overline{1,N})$ are given smooth complex-valued functions. 

\subsection{The multi-frequency ansatz}
\label{sect-multi-frequency-ansatz}

Consider now the unperturbed system ($\epsilon=0$) with $v=v(0)$,
$p^{(\Omega_i)}=p^{(\Omega_i)}(0)$.
This case corresponds to
the source motion at a constant speed, and the equation of motion
\eqref{ur_str_3} in the co-moving co-ordinate system is a linear partial
differential equations (PDE) with
constant coefficients. Usually we expect that due to the wave radiation to
infinity, for large times the response
of a distributed system subjected to a point excitation transforms into a sum
of harmonics with the different frequencies~$\Omega_i$:
\begin{equation}
\mathscr U=
\sum_{i=1}^N 
\mathscr U^{(\Omega_i)}
\equiv
\sum_{i=1}^N 
\hat{\mathscr W}^{(\Omega_i)} 
\exp(-\I\Omega_i\tau
)+\mathrm{c.c.}+o(1), 
\quad
\tau \to \infty.
\label{responce-false}
\end{equation}
The solution in the form of 
Eq.~\eqref{responce-false} is a pure forced oscillation, 
and, therefore, constants 
$\hat{\mathscr W}^{(\Omega_i)}$ 
can be easily found basing on, for example, the stationary
formulation for the problem, or by applying the method of stationary phase
(see \ref{App-with}). 
However, generally Eq.~\eqref{responce-false}
is not true for the system under
consideration. Under certain conditions (in particular, in the case $M>0$,
$K=0$ \cite{gavrilov2002etm}) there is a special frequency
\hbox{$0<\Omega_0<\Omega_\ast$}, 
\begin{equation}
\Omega_\ast\=\sqrt{1-\upsilon^2},
\label{cut-off}
\end{equation}
to which a trapped mode of
oscillation corresponds. This mode is localized near the discrete oscillator.
The frequency defined in 
Eq.~\eqref{cut-off}
is so-called cut-off (or boundary) frequency, discussed in
Appendix~\ref{aux-dispersion}. 
{According to dispersion relation~\eqref{disp-relation}
free waves with frequencies upper
than the cut-off frequency are sinusoidal propagating waves, whereas free waves 
with frequencies below than the cut-off frequency are growing inhomogeneous
waves, which cannot exist if we require boundedness}. 
The expression for $\Omega_0$, necessary and sufficient conditions for the existence of
the trapped mode for the system under consideration are 
obtained in Appendix~\ref{spectral-joined}. If
the trapped mode exists, then 
in the non-resonant case wherein 
\begin{equation}
\Omega_i\not\simeq\Omega_0 \quad  \text{for}
\quad{i=\overline{1,N}}
\label{non-resonant0}
\end{equation}
instead of Eq.~\eqref{responce-false} we get
\begin{equation}
\mathscr U=
\sum_{i=0}^N 
\mathscr U^{(\Omega_i)}
\equiv
\sum_{i=0}^N 
\hat{\mathscr W}^{(\Omega_i)} 
\exp(-\I\Omega_i\tau
)+\mathrm{c.c.}+o(1), 
\quad
\tau \to \infty.
\label{responce-true}
\end{equation}
The additional term 
$
\mathscr U^{(\Omega_0)}
$
corresponds to ``natural'' localized oscillation. 
Constant
$\hat{\mathscr W}^{(\Omega_0)}$ 
(the corresponding amplitude)
can be found using the method
of stationary phase 
(see Appendix~\ref{neodnor}).
Thus, in the case $\epsilon=0$, the response
of the system under consideration is the superposition of a mode of natural oscillation
with frequency $\Omega_0$
and of $N$ modes of forced oscillation, like in the case of a single degree of
freedom system. Note that for the problem under consideration Eq.~\eqref{responce-true}
in the explicit form is formula \eqref{itog}.

Now consider the case of a non-uniform oscillator motion ($\epsilon>0$). In
the non-resonant case (where 
Eqs.~\eqref{non-resonant}, 
\eqref{non-resonant0} are fulfilled for all $T\geq0$) we expect that for large
times the
displacement under the moving oscillator can be approximately found in the
form of the following multi-frequency ansatz:
\begin{gather}
\mathscr U
\simeq
\sum_{i=0}^N 
\mathscr U^{(\Omega_i)}
\equiv
\sum_{i=0}^N 
\mathscr W^{(\Omega_i)}
(T)
\exp\left(-\I\int_0^\tau\Omega_i(T)\,\d T
\right)+\mathrm{c.c.},
\quad
\tau\to\infty,
\label{multifreq-a}
\end{gather}
where amplitudes $\mathscr W^{(\Omega_i)}$ can be represented by the asymptotic
expansions
\begin{gather}
\mathscr W^{(\Omega_i)}(T)=\sum_{j=0}^\infty
\epsilon^j
\mathscr W_j^{(\Omega_i)}(T).
\end{gather}
Additionally, we will require that 
\begin{equation}
\lim_{T\to+0}
\mathscr W^{(\Omega_i)}
(T)=\hat{\mathscr W}^{(\Omega_i)}.
\end{equation}
The quantity $\Omega_0(T)$ in Eq.~\eqref{multifreq-a}
is an immediate value of the trapped mode frequency in the system with fixed
$v=v(T)$, where $T$ should be considered as a time-like parameter.

The applicability of ansatz 
\eqref{multifreq-a}
as a reasonable approximation for the solution of the problem 
under consideration is our hypothesis, which should be verified by numerical
calculations. This hypothesis is supported by several circumstances, namely:
\begin{itemize} 
\item For $\epsilon\to0$ ansatz \eqref{multifreq-a} transforms into
non-stationary solution~\eqref{responce-true}, which can be formally
obtained by the method of stationary phase;
\item
The characteristic time
after which the approximation 
\eqref{responce-true} becomes practically applicable does not depend on
$\epsilon$, and, therefore, is 
$O(1)$, whereas the characteristic time of change for the velocity $v(T)$ and
the external force
amplitudes $p^{(\Omega_i)}(T)$ ($i=\overline{1,N}$) is $O(\epsilon^{-1})$;
\item An asymptotic solution for a
single degree of freedom system with time-varying coefficients has the form of 
Eq.~\eqref{multifreq-a}
\cite{feschenko1967eng,nayfehperturbation}.
\end{itemize}
However, since the problem under consideration is formulated for a PDE with
time-varying coefficients,
to prove the applicability of multi-frequency ansatz 
\eqref{multifreq-a}
in a formal way is a really hard problem. In this sense, our approach is not a pure
asymptotic one, even though 
accepting \eqref{multifreq-a} has a deep asymptotic motivation.
Nevertheless, in what follows, we look for the principal terms
of the amplitudes $\mathscr W_0^{(\Omega_i)}(T)$ using a rigorous asymptotic procedure of
the method of multiple scales. 
We use an approach 
\cite{gavrilov2002etm}
{based 
on the modification of the method of multiple scales}
(Sect.~7.1.6 of \cite{nayfehperturbation})
 for ordinary differential equations (ODE) with slowly varying coefficients.
The corresponding rigorous proofs, which validate such asymptotic approach in
the case of a one degree of freedom system, can be found in
\cite{feschenko1967eng}.
However, since we look for the solution of
a PDE, we need to continue the 
multi-frequency ansatz \eqref{multifreq-a} to a neighbourhood  of the point
$\xi=0$:
\begin{gather}
u(\xi,\tau)=\sum_{i=0}^N u^{(\Omega_i)},
\\
u^{(\Omega_i)}(\xi,\tau)\Big|_{\xi=0}
=
\mathscr U^{(\Omega_i)}(\tau).
\end{gather}

\subsection{The single-frequency ansatz}
\label{sect-single-frequency-ansatz}
We assume that every amplitude $p^{(\Omega_i)}(T)$ is presented in the form of an
asymptotic series:
\begin{equation}
p(T)=\sum_{j=0}^\infty \epsilon ^j p_j(T).
\end{equation}
Here we have dropped the superscript
$(\Omega_i)$ near quantities $p$ and $p_j$ for the aim of simplicity.

We represent the continuation $u^{(\Omega_i)}$ of every term 
$\mathscr U^{(\Omega_i)}$
in the right-hand side of 
\eqref{multifreq-a}
to $\xi\lessgtr0$ as the following single-frequency ansatz:
\begin{gather}
u(\xi,\tau)= W(X,T)\exp
\phi(\xi,\tau),
\label{ansatz-u}
\end{gather}
where 
\begin{gather}
X=\epsilon \xi
\end{gather}
is the slow spatial co-ordinate; $\phi(\xi,\tau)$ such that
\begin{gather}
\phi' = \I\omega(X,T), \qquad \dot{\phi}=-\I\varOmega(X,T),
\label{OSC-fast-phases}
\\
\lim_{X\to\pm0}\varOmega=\Omega_i
\label{lim-Omega}
\end{gather}
is the phase;
\begin{gather}
W(X,T)=\sum\limits_{j=0}^\infty \epsilon^j W_j(X,T)
\label{expansion-basic}
\end{gather}
such that
\begin{gather}
\lim_{X\to\pm0} W(X,T)=\mathscr W(T),
\\
\lim_{X\to\pm0} W_j(X,T)=\mathscr W_j(T)
\label{lim-W}
\end{gather}
is the amplitude. 
The wave-number $\omega(X,T)$ and the frequency $\varOmega(X,T)$
should satisfy dispersion relation 
\eqref{disp-relation}
 and equation 
\begin{equation}
{\varOmega}'_X+{\omega}'_T=0
\label{OSC-dxx-SPRING}
\end{equation}
that follows from
\eqref{OSC-fast-phases}
for all $X$ and $T$ in a neighbourhood of $X=0$.
In this case, the phase $\varphi(\xi,\tauu)$ can be defined by the formula 
\begin{equation}
 \varphi=\I\int(\omega\,\d\xi-\varOmega\,\d\tauu).
\end{equation}
Additionally, we require that 
\begin{gather}
[W]=0,\qquad 
[\varphi]=0.
\label{jumps-Wphi}
\end{gather}

\begin{remark}  
Representations
\eqref{ansatz-u}--\eqref{jumps-Wphi}
are valid and different for all $N+1$ single modes 
$u^{(\Omega_i)}$, i.e.,\ we again have dropped in those equations the superscript
$(\Omega_i)$ near quantities $u,\ W,\ W_j,\ \phi,\ \varOmega,\ \omega$ for the
aim of simplicity. Moreover,  the analytic expressions for
these quantities are generally different for $X\lessgtr0$ since we
additionally require that $u$ satisfy some boundary conditions at infinity
($X\to\pm\infty$). These are vanishing boundary conditions {for
$|\Omega_i|<\Omega_\ast$
(where the wave-numbers are imaginary) and some radiation conditions for
$|\Omega_i|>\Omega_\ast$}
(where the wave-numbers are real). To satisfy these boundary conditions we need to
choose 
for $X\lessgtr0$
different roots~\eqref{wavenumber-expl}.
\label{remark-single}
\end{remark}

{The aim of the analytic work in this paper is to find the principal term of
multi-frequency ansatz \eqref{multifreq-a}. We evaluate independently every
term of \eqref{multifreq-a}, 
which is represented in the form of a single-frequency ansatz 
\eqref{ansatz-u}--\eqref{jumps-Wphi}. 
In what follows, we will get the corresponding solution in
the case when initially the trapped mode in the corresponding system with 
constant $v=v(0)$ 
exists, i.e., if
Eq.~\eqref{v-Kgt0}
or
\eqref{v-Kless0} is fulfilled for $T=0$.
Also, we will demonstrate that the principal term of multi-frequency ansatz 
\eqref{multifreq-a} and the corresponding numerical solution are in excellent
agreement unless the trapped mode disappears at a certain $T$.
We also demonstrate that 
multi-frequency ansatz \eqref{multifreq-a}  is not practically applicable 
after this instant. The
latter case is beyond the scope of the analytic work in this paper.

Finally, we indicate that the principal zero-order terms for all modes
of multi-frequency ansatz 
\eqref{multifreq-a}
$\mathscr U^{(\Omega_i)},\
i=\overline{1,N}$, which correspond
to a forced oscillation, can be found without consideration of their
continuations $u^{(\Omega_i)}$ 
(these terms can be found from equations of zero order approximation).
Looking for the pricipal zero-order term only, we really need 
to introduce the continuation $\mathscr W_0^{(\Omega_0)}(X,T)$
only to calculate the evolution of
amplitude for the trapped mode $\mathscr U_0^{(\Omega_0)}(T)$. 
In the latter case the equations of the first order approximation are necessary.
The corresponding details are given in what follows (see
Sect.~\ref{sec-asymptotic}).


\subsection{The representation for the differential operators}
\label{sec-diff-op}
According to the method of multiple scales 
\cite{nayfehperturbation},
the slow variables $X,\ T,$ and the fast phase $\varphi $ are assumed to be
independent variables. In this way, we represent the differential operators with respect to time and
the spatial co-ordinate in the following form:
\begin{equation}
\label{proizv}
\begin{gathered}
\dot{(\cdot)}=-\I \varOmega \partial_\phi + \epsilon \partial_T, \qquad 
(\cdot)'=\I\omega \partial_\phi +\epsilon \partial_X, \\
\ddot{(\cdot)}=-\varOmega^2 \partial_{\phi\phi}^2 - 2\epsilon \I \varOmega \partial_{\phi T}^2
   -\epsilon \I \varOmega'_T \partial_\phi +O(\epsilon^2), \\
(\cdot)'' = -\omega^2 \partial_{\phi\phi}^2 +2\epsilon \I \omega \partial_{\phi X}^2 
   +\epsilon \I \omega'_X \partial_\phi +O(\epsilon^2),\\
(\cdot)\dot{{\;}}'= \omega \varOmega \partial^2_{\phi \phi}-\epsilon \I
   \varOmega \partial_{\phi X}^2- \epsilon \I \varOmega'_X \partial_\phi
   +\epsilon \I \omega \partial^2_{\phi T}+O(\epsilon^2).
\end{gathered}
\end{equation}

\section{Asymptotic solution}
\label{sec-asymptotic}

\subsection{Contribution from the trapped mode} 
\label{sec-c-trapped}

Accepting of the representation for the contribution from the frequency of
the localized oscillation in the form of the single-frequency
ansatz~\eqref{ansatz-u}--\eqref{jumps-Wphi} (wherein $i=0$ and the superscript
$(\Omega_0)$ is assumed near the corresponding quantities, see Remark~\ref{remark-single})
implies that
\begin{itemize}	
\item Frequency equation 
\eqref{ur_chastot}
for the trapped mode holds for all $T$;
\item Dispersion relation 
\eqref{disp-relation}
 at $\xi=\pm0$
holds for all $T$.
\end{itemize}

At first, we substitute Eq.~\eqref{ansatz-u} 
and representations for differential operators~\eqref{proizv} into the second
Hugoniot condition \eqref{usl_Gu_2}, wherein 
$P(\tau)$ in the right-hand side is expressed by Eq.~\eqref{ur_osc_2}.
Since the representation for the solution at $\xi=0$ in the form of 
multi-frequency ansatz 
\eqref{responce-true}
becomes valid 
after a certain time, whereas $\hat p$ is a finite or exponentially vanishing
function, we take $p=0$ here.
In this way, taking into account 
\eqref{lim-Omega},
one can obtain
\begin{equation}
[\I\omega W  + \epsilon W{}'_X ]  + O(\epsilon^2)= \frac{M\big(-\Omega_0^2 W(0,T)  -2 \nu
\epsilon \I\Omega_0 W(0,T)'_{T} - \nu \epsilon \I\Omega_0{}'_T W(0,T)\big)+K
W(0,T)}{1-\upsilon^2}.
\label{jump-ansatz}
\end{equation}
Now we substitute expansion~\eqref{expansion-basic}
 into Eq.~\eqref{jump-ansatz} and equate
coefficients of like powers $\epsilon$. Taking into account frequency
equation~\eqref{ur_chastot} for $\Omega_0$, one can demonstrate that the equation for
the zeroth order approximation is identically satisfied.
For the first order approximation one can derive:
\begin{equation}
\label{20}
[ W_0{}'_X ] = \frac{-2\nu \I M\Omega_0 W_{0}(0,T)'_{T} - \nu \I M\Omega_0{}'_T
W_0(0,T) }{1-\upsilon^2}.
\end{equation}
Note that Eq.~\eqref{20} does not involve terms, which depend on $W_1$, since
the common multiplier before all such terms equals zero according to the
frequency equation~\eqref{ur_chastot}.

On the other hand, we can define the quantity in the left-hand side of
Eq.~\eqref{20} by  consideration of Eq.~\eqref{ur_str_3}, 
wherein the right-hand side is put to zero,  in the case $\xi\lessgtr0$, and, in particular at $\xi\to\pm0$. To do this, we substitute  ansatz 
\eqref{ansatz-u}--\eqref{jumps-Wphi} and representations \eqref{proizv} 
into Eq.~\eqref{ur_str_3}  and equate
coefficients of like powers $\epsilon$. Taking into account dispersion
relation~\eqref{disp-relation},
 for the zeroth order approximation one
can again demonstrate that the corresponding equation is identically satisfied.
For the first order approximation we obtain:
\begin{multline}
\left((1-\upsilon^2)2\omega -2\upsilon \nu  \Omega_0  \right){W_0}{}'_X\\+
(2\upsilon \omega +2\nu  \Omega_0){W_0}{}'_T + \left((1-\upsilon^2)\omega{}'_X +2
\upsilon \omega{}'_T +\nu  \Omega_0{}'_T+ \omega a \right)W_0=0
\end{multline}
or
\begin{equation}
{W_0}{}'_X= -\frac{(2\upsilon \omega +2\nu  \Omega_0){W_0}{}'_T +
\big((1-\upsilon^2)\omega{}'_X +2 \upsilon \omega{}'_T +\nu  \Omega_0{}'_T+ \omega a
\big)W_0}{(1-\upsilon^2)2\omega -2\upsilon \nu  \Omega_0}
\end{equation}
at $\xi=\pm 0$. Due to Eq.~\eqref{OSC-dxx-SPRING}
one has
\begin{equation}
 {\omega}{}'_X={\omega}{}'_\Omega\,\varOmega{}'_X=
 -{\omega}{}'_\Omega\,{\omega}{}'_T.
\label{OSC-omega-x}
\end{equation}
Using this formula, we can write down:
\begin{equation}
{W_0}{}'_X= -\frac{(2\upsilon \omega +2\nu  \Omega_0){W_0}{}'_T +
\big(-(1-\upsilon^2)\omega{}'_\Omega \omega{}'_T +2 \upsilon \omega{}'_T +\nu
\Omega_0{}'_T+ \omega a \big)W_0}{(1-\upsilon^2)2\omega -2\upsilon \nu
\Omega_0}.
\end{equation}
Here ${\omega}{}'_\Omega$ and ${\omega}{}'_T$ should be calculated in accordance
with Eq.~\eqref{wavenumber-expl}.


Thus, one can obtain
\begin{equation}
\label{W_0X}
    [W_0{}'_X] = - \frac{(\Lambda_0+\Lambda_2)W_0(0,T)+\Lambda_1
    W_0(0,T){}'_T}{(1-\upsilon^2)\I S(\Omega_0)},
\end{equation}
where
\begin{equation}
\begin{gathered}
\Lambda_0\=aB(\Omega_0),\\
\Lambda_1\=2\upsilon B(\Omega_0) +2\Omega_0,\\
\Lambda_2\=
-(1-\upsilon^2)
\big(B{}'_\Omega(\Omega_0)B{}'_T(\Omega_0)-
S{}'_\Omega(\Omega_0)S{}'_T(\Omega_0)\big)
+2\upsilon
B{}'_T(\Omega_0)+\Omega_0{}'_T\\ 
 = (1-\upsilon^2)S{}'_\Omega S{}'_T+\upsilon B{}'_T +\Omega{}'_T.
\end{gathered}
\label{lambdas}
\end{equation}
Now, equating the right-hand sides of Eqs.~\eqref{20} 
and \eqref{W_0X} results in the first approximation equation for
$\mathscr W_0(T) \equiv W_0(0,T)$:
%
\begin{equation}
     \frac{-2 M\Omega_0 \mathscr W_0{}'_T -  M\Omega_0{}'_T \mathscr W_0 }{(1-\upsilon^2)}
     = \frac{(\Lambda_0+\Lambda_2) \mathscr W_0+\Lambda_1 \mathscr W_0{}'_T}{
     (1-\upsilon^2)S(\Omega_0)}.
\end{equation}
The above equation can be transformed to the following one:
\begin{equation}
\frac{\mathscr W_0{}'_T}{\mathscr W_0} =
-\frac{\Lambda_0+\Lambda_2+MS{\Omega_0}{}'_T}{\Lambda_1+2M\Omega_0 S}.
\end{equation}
Substituting expressions~\eqref{lambdas} for $\Lambda_0$, $\Lambda_1$,
$\Lambda_2$ into the above equation, we can derive:

\begin{equation}
\frac{\mathscr W_0{}'_T}{\mathscr W_0} = -\frac{1}{2}\frac{(1-\upsilon^2)S{}'_\Omega
S{}'_T+\upsilon B{}'_T +aB+{\Omega_0}{}'_T+MS{\Omega_0}{}'_T}{\upsilon B
+\Omega_0+M\Omega_0 S}.
\end{equation}
One can rewrite this equation as follows:
\begin{equation}
\frac{\mathscr W_0{}'_T}{\mathscr W_0} = -\frac{1}{2}\frac{(1-\upsilon^2)S{}'_\Omega
S{}'_T+\upsilon B{}'_T +aB+{\Omega_0}{}'_T+M{\Omega_0}{}'_T S +M\Omega_0 S{}'_T - M\Omega_0
S{}'_T}{\upsilon B +\Omega_0+M\Omega_0 S}
\end{equation}
or, equivalently:
\begin{equation}
\frac{\mathscr W_0{}'_T}{\mathscr W_0} = -\frac{1}{2}\frac{(1-\upsilon^2)S{}'_\Omega
S{}'_T+(\upsilon B+\Omega_0+M\Omega_0 S){}'_T - M\Omega_0 S{}'_T}{\upsilon B
+\Omega_0+M\Omega_0 S}.
\end{equation}
Thus, we obtain:
\begin{equation}
\label{tut}
\frac{\mathscr W_0{}'_T}{\mathscr W_0} = -\frac{1}{2}\frac{(1-\upsilon^2)S{}'_\Omega  -
M\Omega_0 }{\upsilon B +\Omega_0+M\Omega_0 S}\,S{}'_T
-\frac{1}{2}\frac{(\upsilon B+\Omega_0+M\Omega_0 S){}'_T }{\upsilon B
+\Omega_0+M\Omega_0 S}.
\end{equation}
Taking into account Eqs.~\eqref{vB+O_0},\eqref{S_Omega_v}, one can demonstrate
that
\begin{equation}
\frac{(1-\upsilon^2)S{}'_\Omega  - M\Omega_0 }{\upsilon B +\Omega_0+M\Omega_0
S}=-\frac{1}{S}.
\label{full-diff}
\end{equation}
%
%
Finally, Eq.~\eqref{tut} can be transformed to the following one:
\begin{equation}
\frac{\mathscr W_0{}'_T}{\mathscr W_0} = \frac{1}{2}\frac{S{}'_T}{S}
-\frac{1}{2}\frac{(\upsilon B+\Omega_0+M\Omega_0 S){}'_T }{\upsilon B
+\Omega_0+M\Omega_0 S}.
\label{tut-final}
\end{equation}
Thus, we represent the right-hand side of our equation in the form of a total
differential of the logarithm of a certain function.
The solution of Eq.~\eqref{tut-final} is as follows:
\begin{equation}
\mathscr W_0 = C\sqrt{\frac{S}{\upsilon B +\Omega_0+M\Omega_0 S}} \, ,
\label{mode-evo}
\end{equation}
where $C$ is an arbitrary complex constant.
Taking into account Eqs.~\eqref{vB+O_0},\eqref{S_Omega_v}, one obtains:
\begin{equation}
\mathscr W_0 
= 
C\frac{\left(1-v^2-{\Omega_0}^2 \right)^{1/4}}
{{\Omega_0}^{1/2}\left(1+M\sqrt{1-v^2-{\Omega_0}^2} \right)^{1/2}}
=
C \sqrt{\frac{M \Omega_0^2-K}{\Omega_0(M^2 \Omega_0^2 -KM +2)}}.
\label{amp-evol}
\end{equation}
To obtain the second equality in the last formula we have taken into account 
frequency equation~\eqref{ur_chastot}.
In the particular case $M=0$ Eq.~\eqref{amp-evol} {yields:}
\begin{equation}
\mathscr W_0 = C\frac{\left(1-v^2-{\Omega_0}^2
\right)^{1/4}}{{\Omega_0}^{1/2}}
=C\sqrt{-\frac{K}{2\Omega_0}}\end{equation}
or
\begin{equation}
\mathscr W_0 =\frac{\tilde C}{\sqrt{\Omega_0}},
\label{amp-evol-m=0}
\end{equation}
where $\tilde C=C\sqrt{-K/2}$ is a constant.
\begin{remark}  
{Note that according to
frequency equation 
\eqref{ur_chastot}
$K$ should be negative if the trapped mode exists in the case
$M=0$.}
\end{remark}
\begin{remark}  
\label{R-L-G}
Formula  \eqref{amp-evol-m=0}, which 
describes the evolution for the amplitude of
the trapped mode in the case $M=0$, $K<0$, coincides with the corresponding
formula for a linear oscillator with spring of slowly time-varying stiffness
\cite{nayfehperturbation} ({the Liouville--Green approximation}). The
particular case $M=0$, $K<0$ of the problem under consideration is the only
one known for us system 
\cite{shishkina2018non,arXiv:1805.07382,gavrilov-da70,indeitsev2016evolution,gavrilov2002etm}
with trapped mode for which the corresponding formula
has {this simple classical form}.
\end{remark}
In the particular case $K=0$ Eq.~\eqref{amp-evol} yields:
\begin{equation}
\mathscr W_0 
= 
C\frac{\left(1-v^2-{\Omega_0}^2 \right)^{1/4}}
{{\Omega_0}^{1/2}\left(1+M\sqrt{1-v^2-{\Omega_0}^2} \right)^{1/2}}
=
C \sqrt{\frac{M \Omega_0}{M^2 \Omega_0^2 +2}}.
\label{amp-evol-K=0}
\end{equation}

\subsection{Contribution from the modes of forced oscillation}
\label{sec-c-forced}

Again, we represent every mode of forced oscillation in the form of the single-frequency
ansatz~\eqref{ansatz-u}--\eqref{jumps-Wphi} (wherein $i=\overline{1,N}$ and superscript
$(\Omega_i)$ is assumed near the corresponding quantities, see
Remark~\ref{remark-single}).
In this way, one gets analogously to Eq.~\eqref{jump-ansatz}
\begin{equation}
[\I\omega W ]  + O(\epsilon)=
\frac{(-M\Omega_0^2+K)\, W(0,T)-p}{1-\upsilon^2}.
\label{jump-ansatz-forced}
\end{equation}
Now we substitute expansion~\eqref{expansion-basic}
 into Eq.~\eqref{jump-ansatz-forced} and equate
coefficients of like powers $\epsilon$.
For the zeroth order approximation one obtains a non-trivial equation:
\begin{equation}
[\I\omega ]\, \mathscr W_0=
\frac{(-M\Omega_0^2+K)\,\mathscr W_0-p_0(T)}{1-\upsilon^2}.
\label{jump-ansatz-forced0}
\end{equation}
Resolving the last equation yields 
\begin{equation}
\mathscr W_0=p_0(T)\,\mathscr G(0,\Omega_i),
\end{equation}
where $\mathscr G(0,\Omega_i)$
is the Green function defined by \eqref{Green-functionO-lower} or \eqref{Green-functionO-upper}.

In the particular case, where $\Omega_i=0$ and $p_0=\const$ is the weight,
this yields
\begin{equation}
\mathscr W_0=\frac{p_0}{2\sqrt{1-\upsilon^2}+K}.
\label{U-weight}
\end{equation}


\subsection{Calculation of the unknown constants}
\label{sec-c-constants}
{Taking into account the contributions from the frequency of localized
oscillation $\Omega_0$, from the frequencies of forced oscillation $\Omega_i$},
and the corresponding complex conjugate terms, {one obtains}
\begin{gather}
\mathscr U=\sum_{i=0}^N \mathscr{U}^{(\Omega_i)}+\mathrm{c.c.} + O(\epsilon)=
\mathscr U_{\mathrm{forced}}+\mathscr U_{\mathrm{free}}
+ O(\epsilon),
\label{U_ne_0-1}
\\
\begin{multlined}       
\mathscr{U}_{\mathrm{forced}} 
\= \sum_{i=1}^N \mathscr{U}^{(\Omega_i)} + \mathrm{c.c.}=
2\sum_{i=1}^N\left|{p^{(\Omega_i)}}(T)\, \mathscr G\big(0,\Omega_i(T)\big)\right|
\\\qquad\times\cos\left(\int_0^\tau\Omega_i(T)\,\d T-\arg\Big(
{p^{(\Omega_i)}}(T)\, \mathscr G\big(0,\Omega_i(T)\big)
\Big)\right),
\end{multlined}
\label{U_ne_0-1-forced}
\\
\mathscr U_{\mathrm{free}}
\= \mathscr{U}^{(\Omega_0)} + \mathrm{c.c.}=
C_0\,
\frac{\left(1-v^2(T)-{\Omega_0}^2(T) \right)^{1/4}
\sin \left(\int_0^\tau \Omega_0(T)\, \d T - D_0 \right)
}
{{\Omega_0}^{1/2}(T)\left(1+M\sqrt{\strut 1-v^2(T)-{\Omega_0}^2(T)} \right)^{1/2}}
.
\label{U_ne_0-1-free}
\end{gather}
To find the unknown real constants $C_0$ and $D_0$ related to the complex
constant $C$ (introduced in~\eqref{mode-evo}) as
\begin{equation}
C_0=2|C|,\qquad D_0=\arg C,      
\end{equation}
one should 
match the right-hand
sides of Eq.~\eqref{U_ne_0-1} taken at $T=0$ 
and Eq.~\eqref{itog}. 
In this way the right-hand
side of Eq.~\eqref{U_ne_0-1-forced}
transforms into the first term in the right-hand side of Eq.~\eqref{itog}.
Equating the second terms yields
\begin{gather}
C_0= \frac{\big ( 1-\upsilon^2(0)-\Omega_0^2(0)\big)^{1/4}\,
\big|\mathscr{F}\{p(\tau,0)\}\big(\Omega_0(0)\big)\big|}
{{\Omega_0^{1/2}(0)} \left(1+M \sqrt{1-\upsilon^2(0)-\Omega_0^2(0)}\right)^{1/2}}
\label{C0-def}
,
\\
D_0=\arg \mathscr{F}\{p(\tau,0)\}\big(\Omega_0(0)\big).
\label{D0-def}
\end{gather}
Here and in what follows $\mathscr F$ is a symbol of the Fourier transform with respect to time
$\tau$,
see the details in \ref{neodnor}.

\subsection{Analytic expression for the unknown internal force}
\label{sect-c-iforce}

Using Eqs.~\eqref{ur_osc_2-dupe}, \eqref{U_ne_0-1}--\eqref{U_ne_0-1-free} {one gets}:
\begin{gather}
\label{P-analytic}
{P} = p+
P_{\mathrm{forced}}+  P_{\mathrm{free}}
+ O(\epsilon),
\\
\begin{multlined}       
P_{\mathrm{forced}}
\= 2\sum_{i=1}^N 
\left|{p^{(\Omega_i)}}(T) \mathscr G\big(0,\Omega_i(T)\big)\right|\,(M\Omega_i^2(T)-K)
\\\qquad\times\cos\left(\int_0^\tau\Omega_i(T)\,\d T-\arg\Big(
{p^{(\Omega_i)}}(T)\, \mathscr G\big(0,\Omega_i(T)\big)
\Big)\right),
\end{multlined}
\label{P-analytic-forced}
\\
P_{\mathrm{free}}
\=
C_0\,(M\Omega_0^2(T)-K)
\frac{\left(1-v^2(T)-{\Omega_0}^2(T) \right)^{1/4}
\sin \left(\int_0^\tau \Omega_0(T)\, \d T - D_0 \right)
}
{{\Omega_0}^{1/2}(T)\left(1+M\sqrt{\strut 1-v^2(T)-{\Omega_0}^2(T)}
\right)^{1/2}}.
\label{P-analytic-free}
\end{gather}

\section{Numerics}
\label{sect-numerics}
\subsection{Integral equation for the unknown internal force}
\label{sec-i-eq}
The solution satisfying Eq.~\eqref{ur_str_2-dim},
and initial conditions~\eqref{initial-cond} can be
written as the convolution over $x$ and $t$ of the fundamental solution of the
Klein-Gordon PDE \eqref{fund-sol}
with the right-hand side
of Eq.~\eqref{ur_str_2-dim}. At the point under the moving inclusion $x=\ell(t)$ 
the solution is as follows \cite{kaplunov1986vibrations,Non-Stationary}:
\begin{multline}
\U(t)=P(t)\delta(x-\ell(t))\,\ast\,\Phi(x,t)\Big|_{x=\ell(t)}\\=
\frac{H(t)}{2}\int _0^t  {H}\left(  1-
\frac{|\ell(t)-\ell(\tau)|}{t-\tau} \right)  P(\tau)J_0 \left(\sqrt{(t-\tau)^2 -
(\ell(t)-\ell(\tau))^2} \right) \d \tau,
\label{U-conv}
\end{multline}
where $J_0(\cdot)$ is the Bessel function of the first kind of zero order, and 
the asterisk is the symbol for the convolution operator.
In the subcritical case 
\eqref{v-is-subcritical}
the Heaviside function in the integrand in the right-hand side of 
\eqref{U-conv}
is equal to one. 

At the same time at the point under the inclusion in the case $M>0$ the same
solution can be found as the convolution over $t$ of the fundamental solution
\eqref{fund-sol-osc}
of ODE describing the linear oscillator with the right-hand side of
Eq.~\eqref{ur_osc_2}:
\begin{equation}
\U(t)=
\big(-P({t}) +p({t})\big)\,\ast\,\Psi(t)
=
H(t)\int_0^t \left(p(\tau) -P(\tau) \right)
\Psi(t-\tau)
\,\d \tau.
\label{U-via-osc}
\end{equation}
Equating the right-hand sides of Eqs.~\eqref{U-conv} and \eqref{U-via-osc}
one gets a Volterra integral equation of the first kind for the unknown internal
force $P(t)$,
which is valid for $t>0$.
Differentiating this equation with respect to $t$ yields the following
Volterra integral
equation of the second kind:
\begin{multline}
P(t)=\int _0^t P(\tau) 
\Bigg(\frac{t-\tau -\big(x(t) - \ell(\tau)\big)\ell'_t(t)}
{\sqrt{(t-\tau)^2 - \big(\ell(t)-\ell(\tau)\big)^2}} 
\,
J_1 \left(\sqrt{(t-\tau)^2 - (\ell(t)-\ell(\tau))^2}
\right)
\\
-2\Psi'_t(t-\tau)
\Bigg)\d \tau 
+ 
2\int _0^t p(\tau) \Psi'_t(t-\tau)
\d \tau,
\label{M>0,Kge0}
\end{multline}
where $J_1(\cdot)$ is the Bessel function of the first kind of the first order.
The last formula is written in the simplified form, which takes into account
that restriction \eqref{v-is-subcritical} is satisfied.

In the special case $M=0$ Eq.~\eqref{ur_osc_2} is not an ODE any more. 
In the latter case, the Volterra integral equation for unknown $P(t)$
can be obtained by substituting of 
Eq.~\eqref{U-conv} into Eq.~\eqref{ur_osc_2}. Provided that 
\eqref{v-is-subcritical} is satisfied,
this yields 
\begin{equation}
P(t) = -\frac{K}{2} \int^t_0 P(\tau) J_0 \left(\sqrt{(t-\tau)^2 -
(\ell(t)-\ell(\tau))^2} \right)\; \mathrm{d} \tau + p(t). 
\label{M=0,K<0}
\end{equation}

The numerical solution of integral equation
\eqref{M>0,Kge0}
or 
\eqref{M=0,K<0} can be compared with the analytic expression for the internal
force $P$ 
\eqref{P-analytic}.

The methodology to solve the obtained integral equation numerically is completely
analogous to the one used in our previous paper
\cite{shishkina2018non}, where it is 
discussed in detail. To obtain the numerical solution for the displacement
$\mathscr U(t)$, 
we use formula \eqref{U-conv} and compute  the convolution of the
numerically obtained internal force $P(t)$ with the fundamental solution of
the Klein-Gordon equation.
\subsection{Comparison between analytic and numerical results}
\label{sec-comparison}

\subsubsection{Pure free oscillation in the case $M>0$, $K>0$}
\label{sec-free-gg}
Considering a pure free localized oscillation we take 
\begin{equation}
 \hat p(\tau,T))=\tau_0^{-1}\big(H(\tau)-H(\tau-\tau_0)\big),\qquad
 p^{(\Omega_i)}\equiv0,
\label{rectangle}
\end{equation}
where $\tau_0$ is a small positive constant. Quantity $\hat p(\tau)$ weakly converges to
$\delta(\tau)$ as $\tau_0\to+0$. In the limiting case we have
\begin{equation}
\mathscr{F}\{p(\tau,0)\}\big(\Omega_0(0)\big)=1.
\label{Fp-delta}
\end{equation}
The last equation should be used to define the unknown constants 
\eqref{C0-def}, \eqref{D0-def}.

As generally, we assume that initially in the system with $v=v(0)$ condition 
\eqref{v-Kgt0} is satisfied and the trapped mode exists, i.e.,\ we can use
the constructed analytic solution.
In Fig.~\ref{Mg0-Kg0.pdf} we compare the results obtained for a monotonically increasing
$v(T)$. We present results for the internal force $P(\tau)$ at sub-plot (a) and
displacement $\mathscr U(\tau)$ at sub-plot (b).
The yellow span in each sub-plot corresponds to the time interval, where the
solution is still sub-critical, but restriction 
\eqref{v-Kgt0} is not satisfied. Thus, our asymptotic solution is defined only
for the first stage of the motion (for the time values to the left of the yellow
span). 
The left boundary of the span 
corresponds to the instant when the immediate value of the localized mode
frequency approaches  the cut-off frequency 
\eqref{cut-off} (this corresponds to the disappearing of the trapped mode).
The right boundary of the span
corresponds to the instant of overcoming the critical speed ($v=1$).
The asymptotic solution approaches the numerical one very quickly.
One can observe that variation of the amplitude of the
localized oscillation is more pronounced in the case of the internal force
$P(\tau)$ than for the displacement $\mathscr U(\tau)$.  The
divergence between the asymptotic and numerical solutions begins again considerable
in a left neighbourhood of the yellow span. At the left boundary of the yellow span
the amplitudes of both analytic solutions $P(\tau)$ and $\mathscr U(\tau)$ become
zero. It is interesting that the amplitude of the internal force $P$ observable for the
numerical solution begins again to grow within the span, whereas 
the corresponding amplitude of the displacement $\mathscr U$ decreases
monotonically within the span. For the time being, the nature of the
characteristic frequency which corresponds to the oscillation within the span
is not absolutely clear for us, but according to our hypothesis the oscillation after
disappearing of the trapped mode 
can be described as a resonant solution describing overcoming the
cut-off frequency.

\begin{figure}[hp]
\centering{\includegraphics[width=0.85\textwidth]{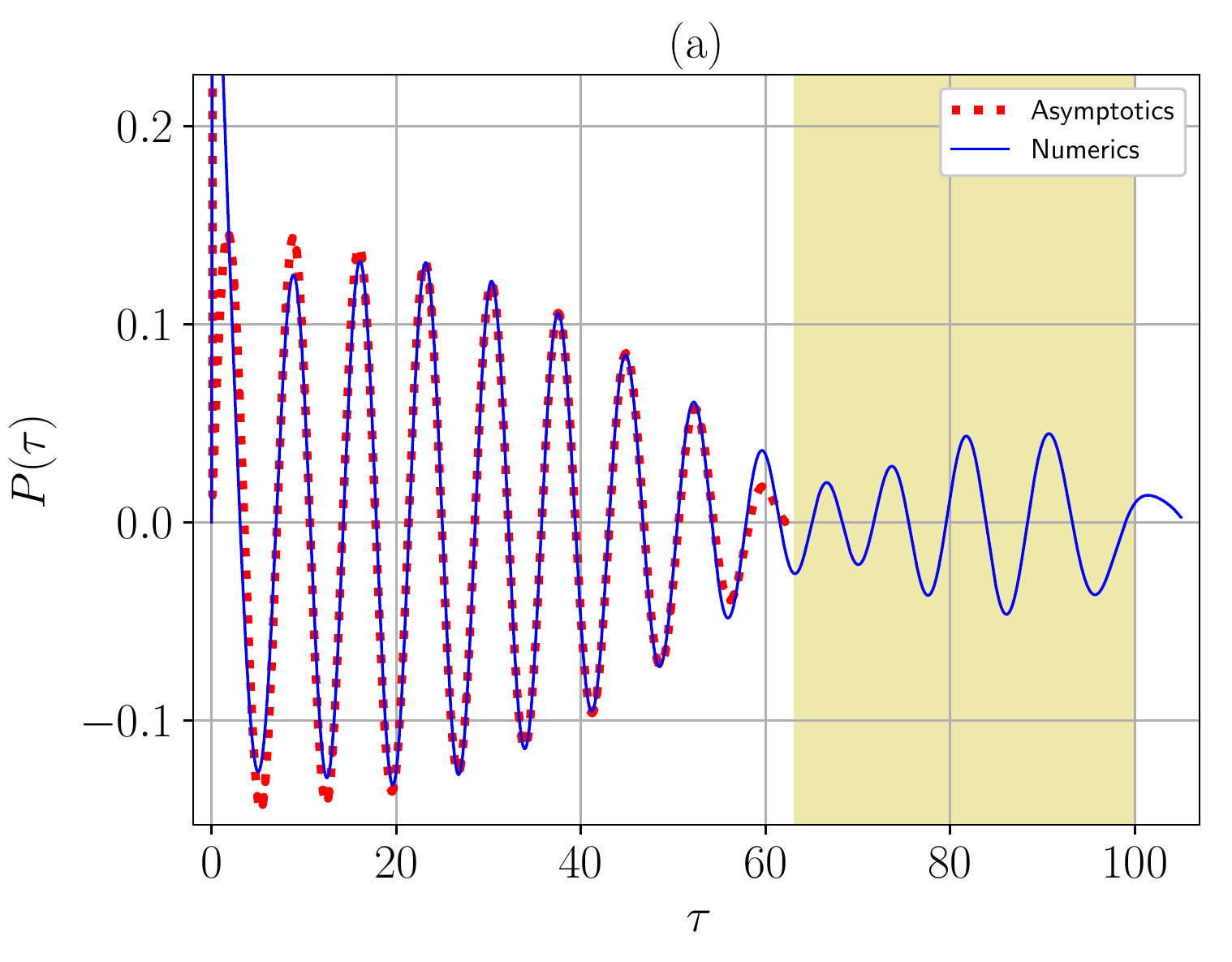}}
\centering{\includegraphics[width=0.85\textwidth]{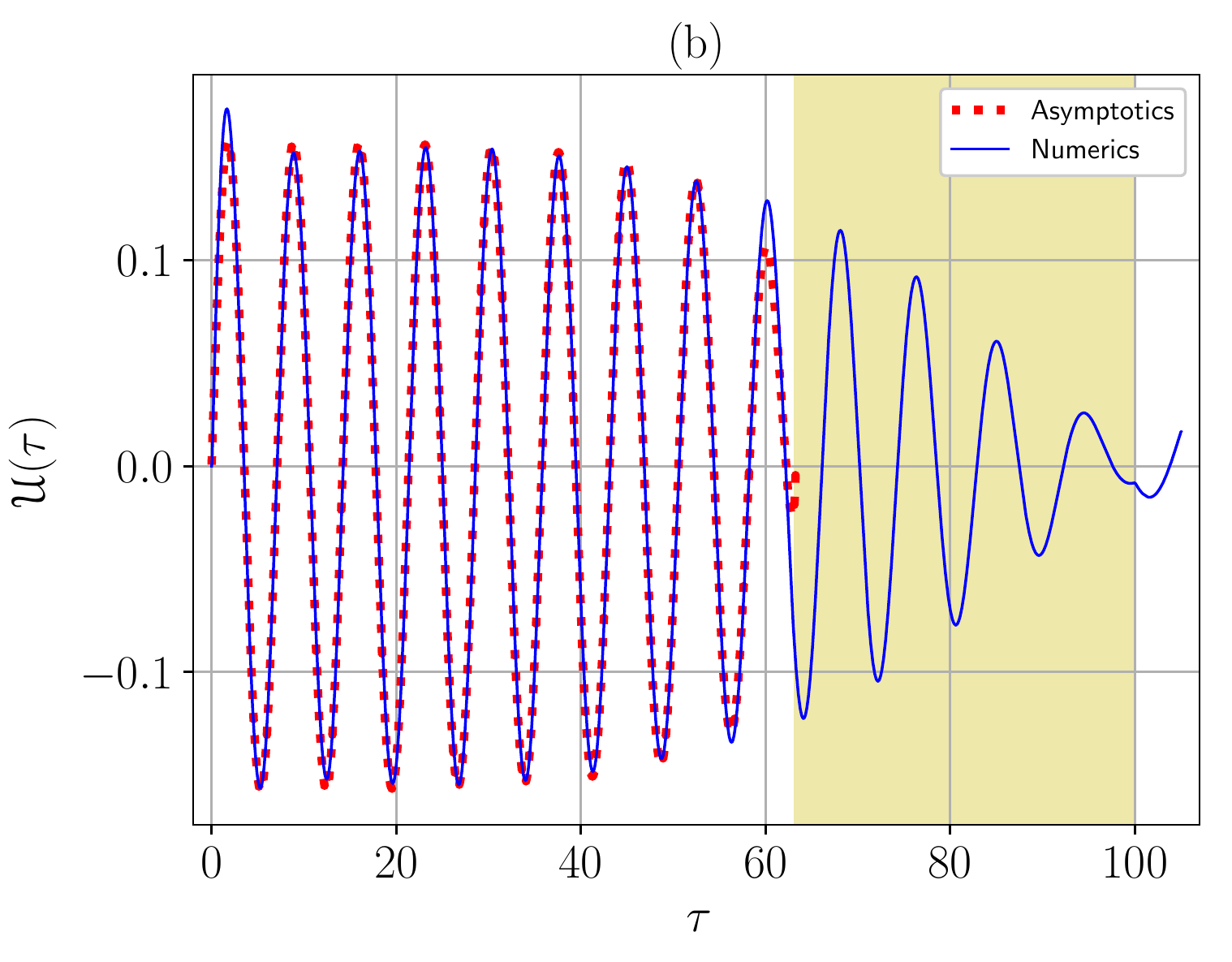}}
\caption{
Comparing the asymptotic solution in the form of 
Eqs.~\eqref{U_ne_0-1}--\eqref{P-analytic-free}
with the corresponding numerical solution for the accelerating oscillator moving at speed
$v=T$: $\epsilon=0.01$, $M=5$, $K=3$. (a) The internal force, (b)
the displacement. The
yellow span in each sub-plot corresponds to the second stage of the motion,
when the solution is still
sub-critical, but restriction \eqref{v-Kgt0} is not satisfied.}
\label{Mg0-Kg0.pdf}
\end{figure}

\afterpage{\clearpage}
\subsubsection{Pure free oscillation in the case $M>0$, $K=0$: comparison with
the results of the previous paper
\cite{gavrilov2002etm}}
\label{sect-comparison-old}
The particular case under consideration was considered in the previous paper 
\cite{gavrilov2002etm}, where the approach based on the method of multiple
scales was suggested at the first time. The final formula describing the
evolution of the amplitude of the localized oscillation was 
\begin{equation}
\mathscr W_0^\mathrm{\,old}=C\sqrt{\frac{1-v^2}{\Omega_0(M^2\Omega_0^2+2)}},
\label{W-old}
\end{equation}
see \cite{gavrilov2002etm}, Eq.~(5.15) in that paper. This formula is in the contradiction
with the result, which we have got in this paper, see Eq.~\eqref{amp-evol-K=0}.
Thus, one of these two formulae is definitely erroneous. The problem is
that the analytic calculations in \cite{gavrilov2002etm} {are extremely complicated
comparing to the current paper, where the technique based on representing of
the right-hand side of the first approximation equation in the form of the
total differential is suggested and applied.} 
Note that previously in \cite{gavrilov2002etm} the corresponding formula was obtained as
a product of ten indefinite integrals.
We were sure that old asymptotics is correct since it describes the
numerical results quite well. Therefore,
analysing the contradiction,
we want to check which asymptotics corresponds to numerical
results better. First, we compare the old asymptotics, the new one and
numerics in the case considered in the plots presented in \cite{gavrilov2002etm},
namely $M=2$, see Fig.~\ref{Mis0.pdf}.%
\begin{figure}[hp]
\centering{\includegraphics[width=0.9\textwidth]{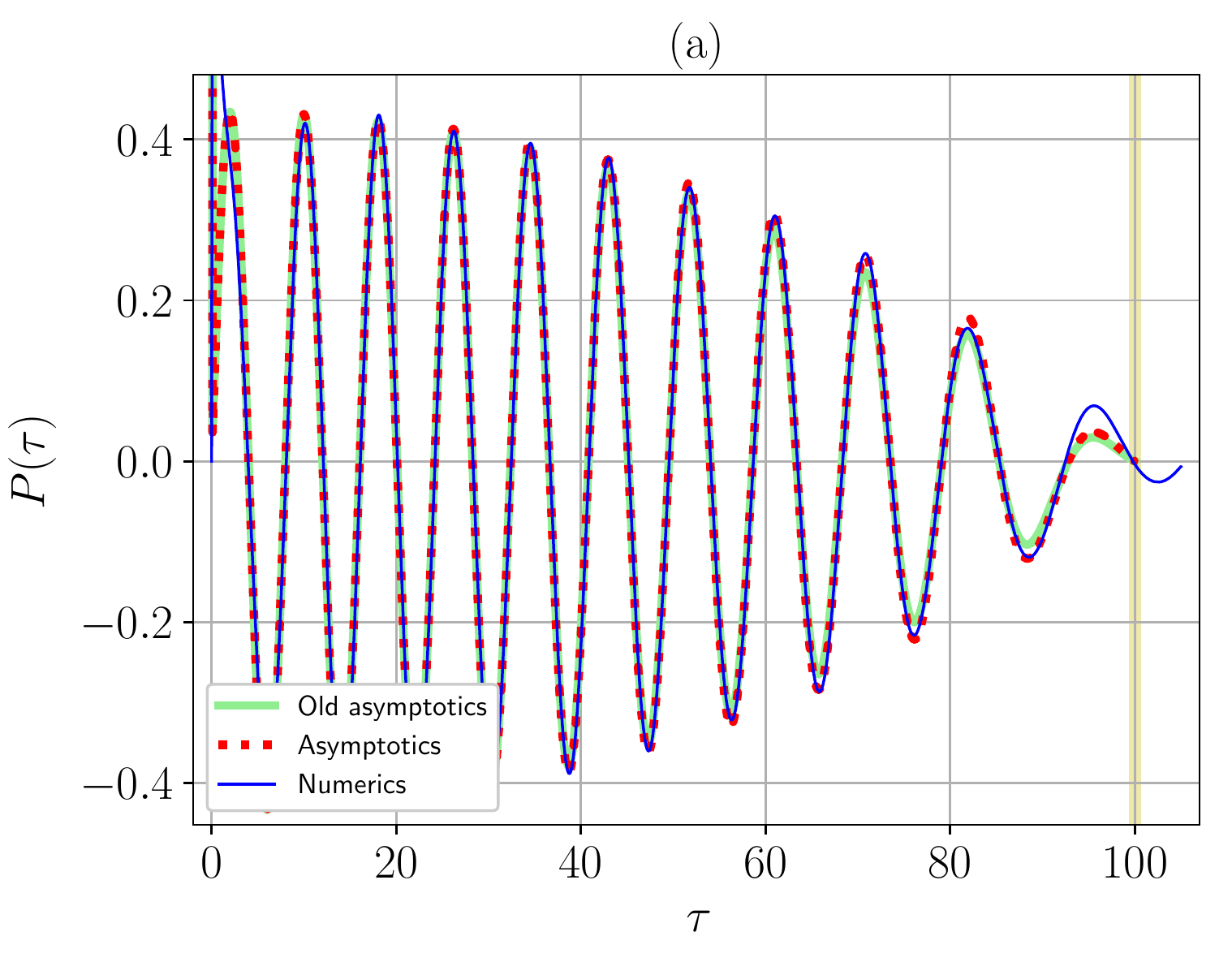}}
\centering{\includegraphics[width=0.9\textwidth]{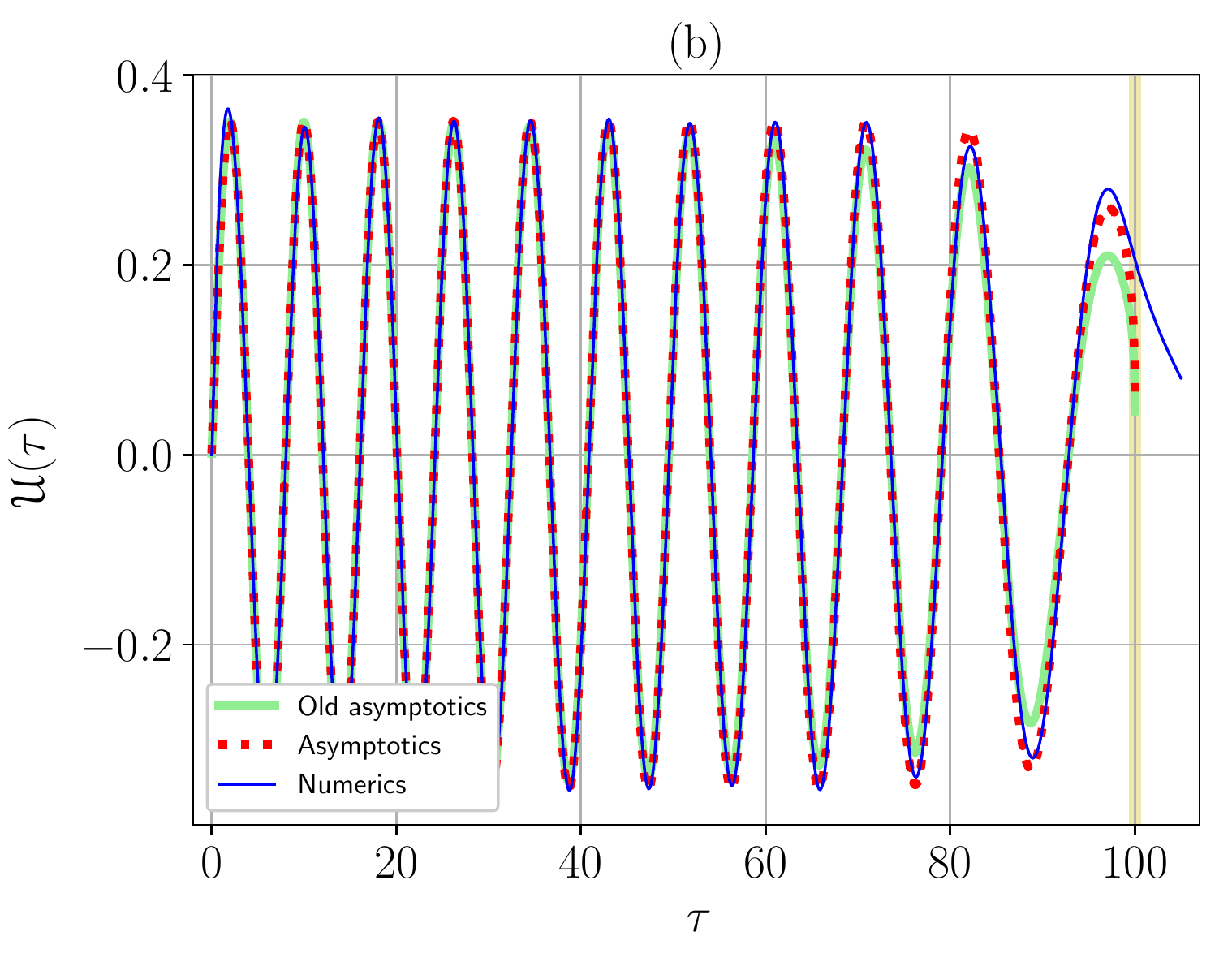}}
\caption{
Comparing the asymptotic solution in the form of 
Eqs.~\eqref{U_ne_0-1}--\eqref{P-analytic-free}, the old asymptotic solution
obtained in \cite{gavrilov2002etm},
and the corresponding numerical solution for the accelerating oscillator
moving at the speed
$v=T$: $\epsilon=0.01$, $M=2$, $K=0$. (a) The internal force, (b)
the displacement. The yellow vertical line in each sub-plot corresponds to the instant
of overcoming the critical speed $v=1$.}
\label{Mis0.pdf}
\end{figure}
\afterpage{\clearpage}
This figure does not allow one to
make the decision, since both asymptotic formulae seem to work well, at
least for times, when trapped mode frequency $\Omega_0$ is far enough from
the cut-off frequency $\Omega_\ast$. To analyse the difference between the
results, we have introduced the ratio $\varkappa$ of the normalized
amplitudes%
\footnote{We have introduced and used the
normalized amplitudes investigating other problem considered in \cite{gavrilov-da70}.}
$
{\mathscr W_0}
/
(\mathscr W_0|_{v=0})
$
and
$
{\mathscr W_0^\mathrm{\,old}}
/
({\mathscr W_0^\mathrm{\,old}}|_{v=0})
$:
\begin{equation}
\varkappa=
\frac{\mathscr W_0}{\mathscr W_0^\mathrm{\,old}}
\left(\left.
\frac{\mathscr W_0}{\mathscr W_0^\mathrm{\,old}}
\right|_{v=0}\right)^{-1}.
\end{equation}
The plots of the coefficient $\varkappa(M)$ for various values of $v$ are
presented in Fig.~\ref{curves.pdf}. One can see that the more mass $M$, the
more ratio $\varkappa$. 
\begin{figure}[hptb]
\centering{\includegraphics[width=0.9\textwidth]{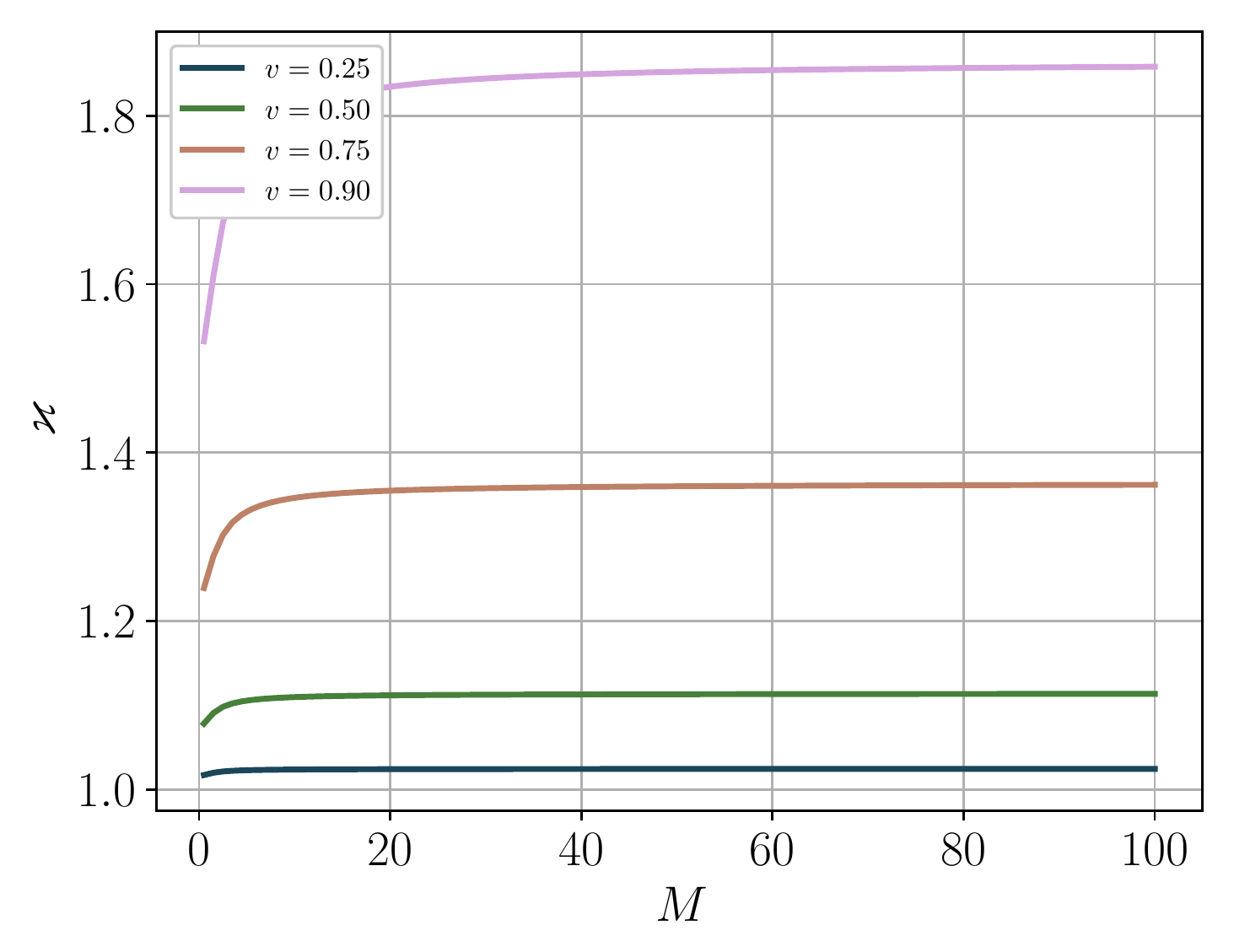}}
\caption{The ratio $\varkappa$ of the normalized
amplitudes
$
{\mathscr W_0}
/
(\mathscr W_0|_{v=0})
$
and
$
{\mathscr W_0^\mathrm{\,old}}
/
({\mathscr W_0^\mathrm{\,old}}|_{v=0})
$ versus $M$ calculated for various values of $v$ in the case $K=0$}
\label{curves.pdf}
\end{figure}
\afterpage{\clearpage}
In Fig.~\ref{Mgg0.pdf} we compare the old asymptotics, the new one and
numerics in the case $M=100$. One can see that the new asymptotics is definitely
better describes numerics, thus, we make a decision that the error is in the old
calculations.
\begin{figure}[hp]
\centering{\includegraphics[width=0.85\textwidth]{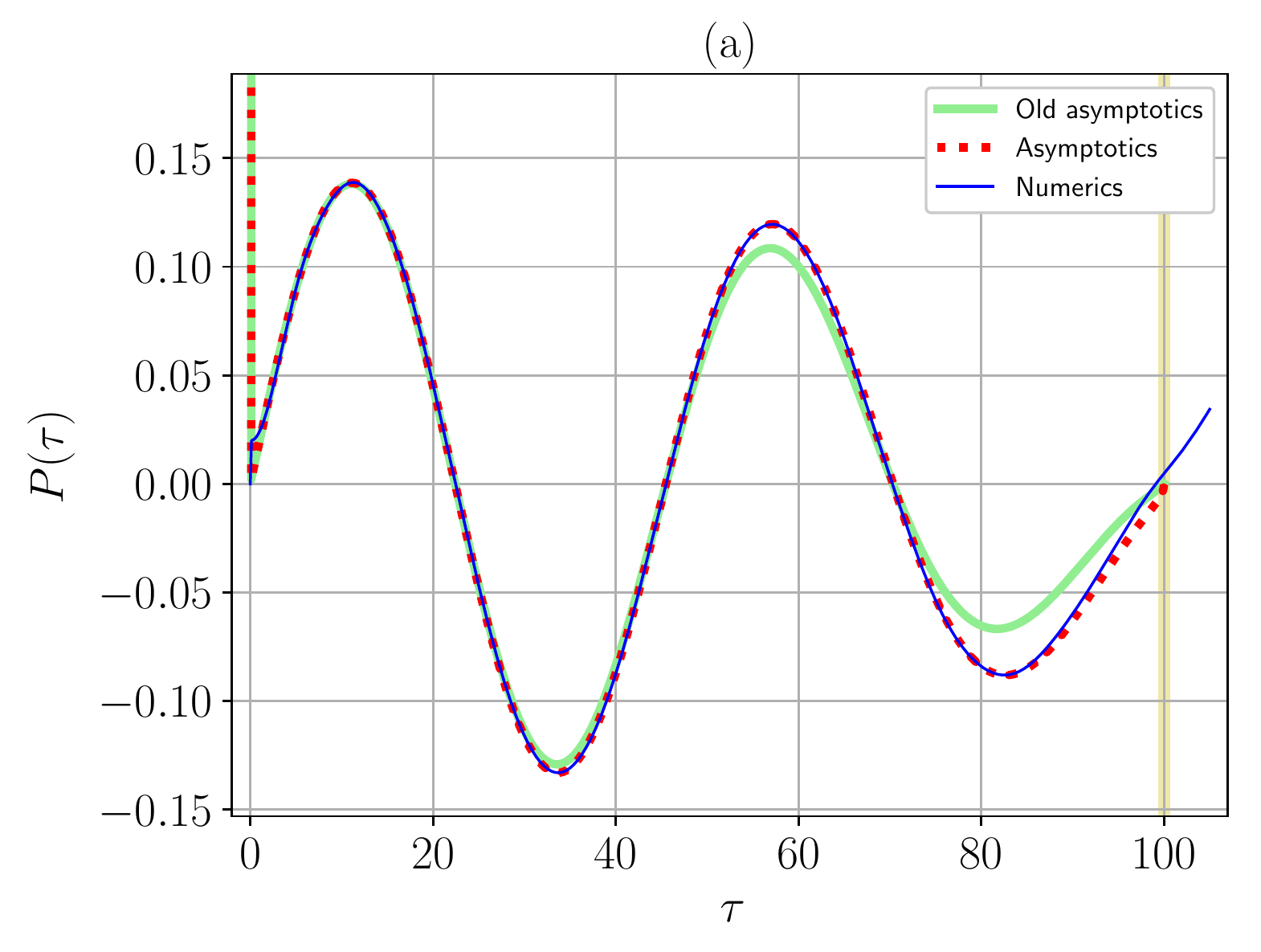}}
\centering{\includegraphics[width=0.85\textwidth]{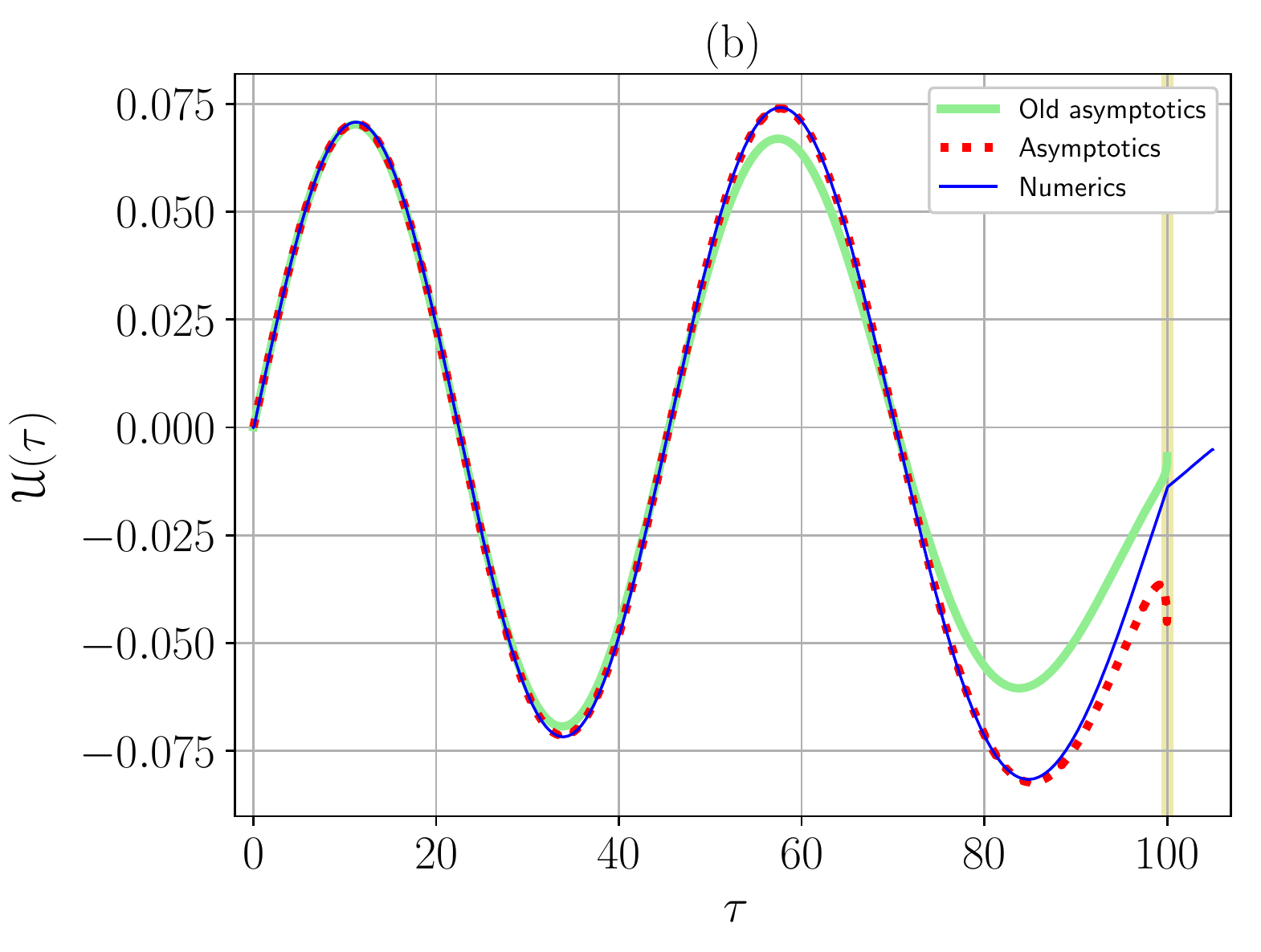}}
\caption{
Comparing the asymptotic solution in the form of 
Eqs.~\eqref{U_ne_0-1}--\eqref{P-analytic-free}, the old asymptotic solution
obtained in \cite{gavrilov2002etm},
and the corresponding numerical solution for the accelerating oscillator
moving at the speed
$v=T$: $\epsilon=0.01$, $M=100$, $K=0$. (a) The internal force, (b)
the displacement. The yellow vertical line in each sub-plot corresponds to the instant
of overcoming the critical speed $v=1$.}
\label{Mgg0.pdf}
\end{figure}
\afterpage{\clearpage}

Finally, we carefully analyse the calculations in \cite{gavrilov2002etm} and
discovered the error, which emerges when calculating quantity $\Phi_3$, see
Eq.~(5.10) in that paper. In \cite{gavrilov2002etm}, to calculate $\Phi_3$, the
frequency equation for the trapped mode differentiated with respect to
$\Omega_0$ was used for the aim of equations simplification.
Since the frequency equation is defined only for the trapped mode
frequency (and it is not valid in the neighbourhood of this frequency), 
this operation is senseless. We also carefully analysed {our subsequent
studies \cite{gavrilov-da70,indeitsev2016evolution,shishkina2018non}}, where
the approach based on the method of multiple scales was
applied to a number of problems, and now we are sure that the same error was
never repeated.

\subsubsection{Pure free oscillation in the case $K<0$}
\label{sec-free-l}
Again, we assume that initially in the system with $v=v(0)$ condition 
\eqref{v-Kless0} is satisfied and the trapped mode exists, i.e.,\ we can use
the constructed analytic solution. In the case $K<0,\ M>0$ the kernel and the free term of integral equation 
\eqref{M>0,Kge0} grow exponentially as $t\to\infty$. This leads to an
oscillatory numerical instability, which is observed after certain value of time
even in the case of a uniform motion (or even for non-moving mass-spring
system), see Fig.~\ref{Unstable.pdf}. Since in the case of a uniform
motion, Eq.~\eqref{itog} is an ``exact'' asymptotics got by the method of stationary
phase, we guess that we observe a {numerical} instability and not a
physical one.
\begin{figure}[htbp]%
\centering{\includegraphics[width=0.85\textwidth]{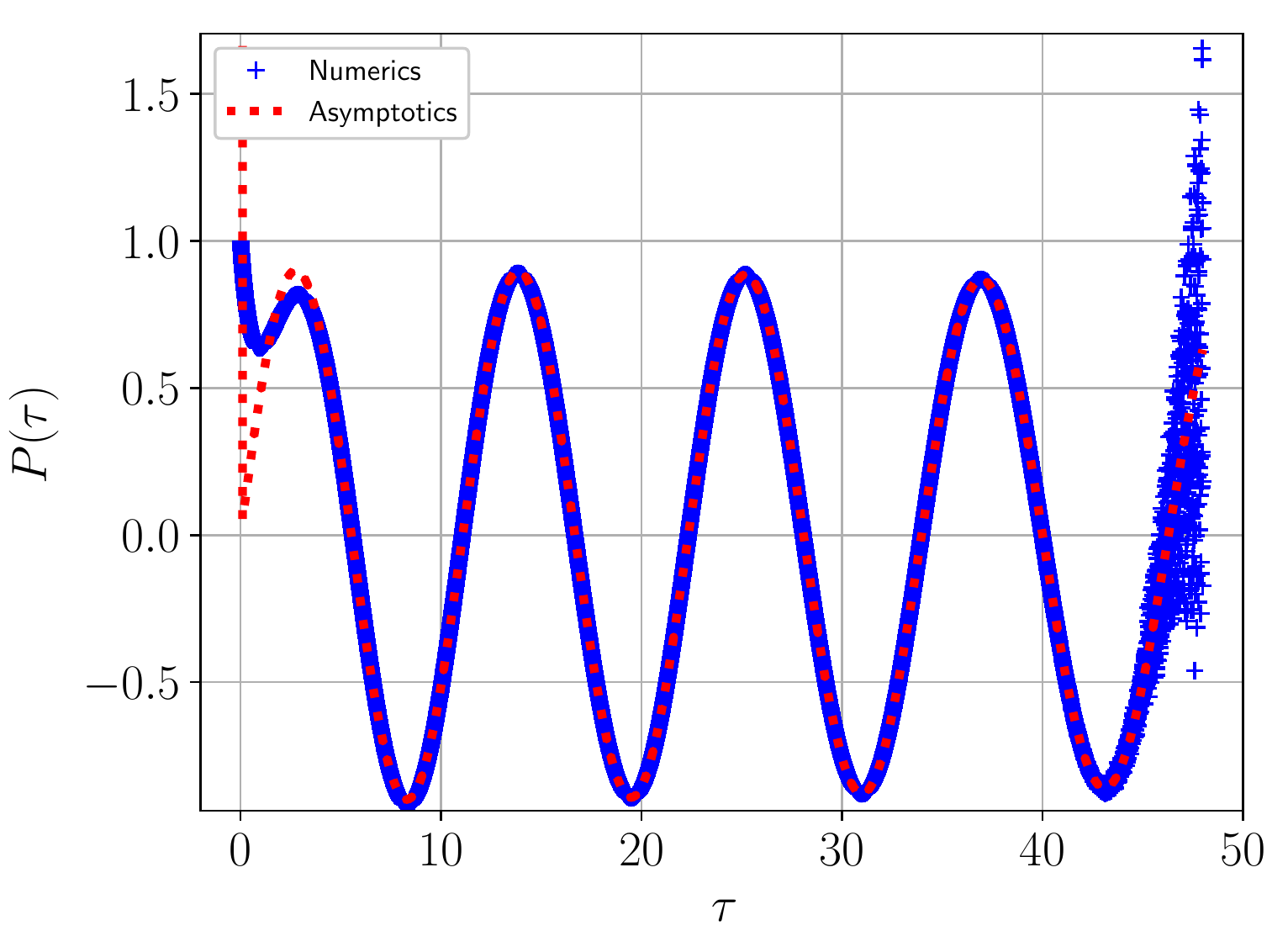}}
\caption{
Numeric instability observed when solving integral equation 
\eqref{M>0,Kge0} in the case $K<0,\ M>0$:
$v=0$, $M=2$, $K=-1$.} 
\label{Unstable.pdf}
\end{figure}

Apparently, we can get rid of this kind of instability using floating-point 
arithmetics with higher precision. Increasing of $M$ leads to the more
moderate growth
of the kernel and the free term. On this way we can get the numerical results
for bigger~$\tau$. Note that in the special case $M=0$ the integral equation
has the form of Eq.~\eqref{M=0,K<0}, which does not involve any exponentially
growing functions; therefore, we do not observe any instability.

\begin{figure}[hp]%
\centering{\includegraphics[width=0.85\textwidth]{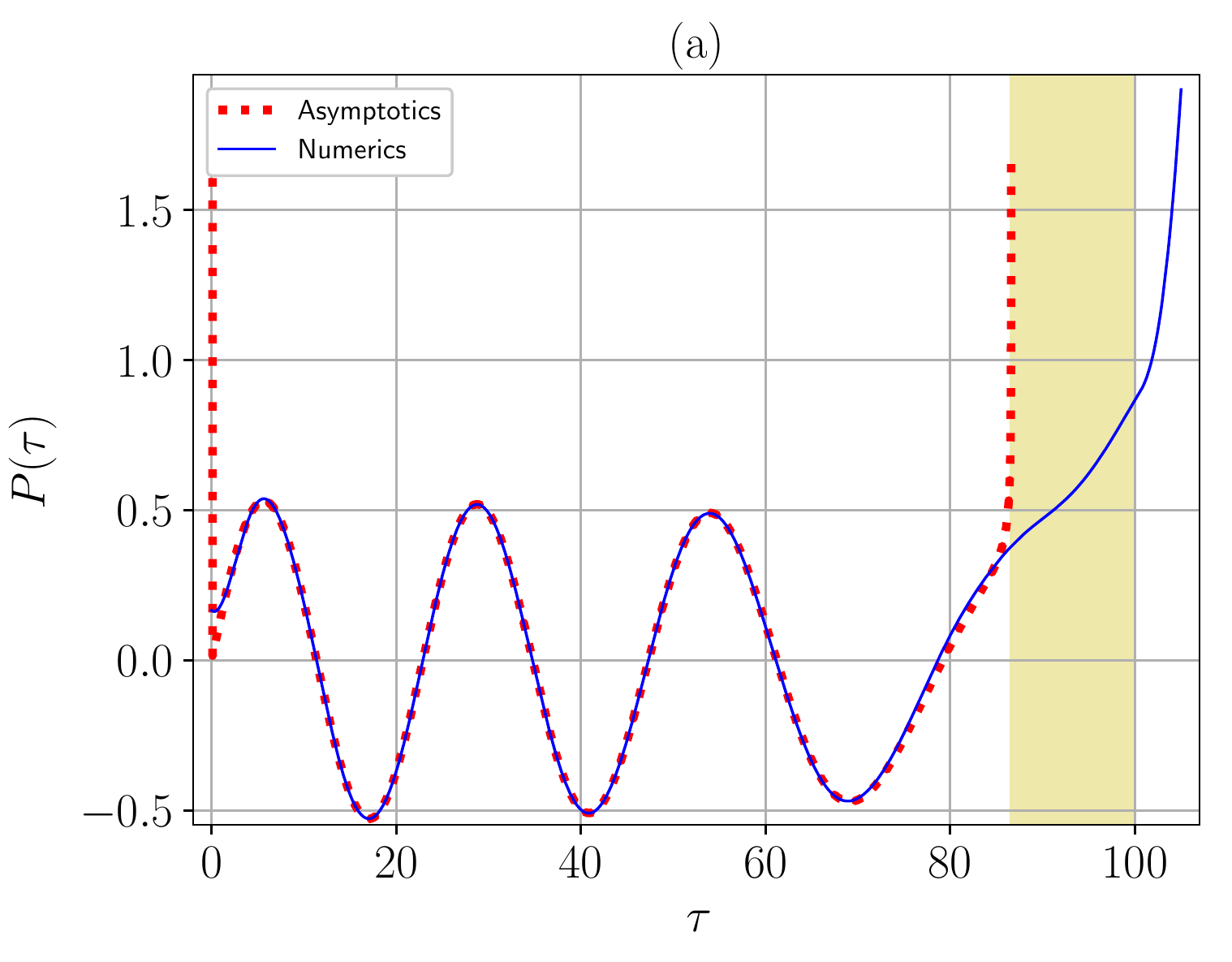}}
\centering{\includegraphics[width=0.85\textwidth]{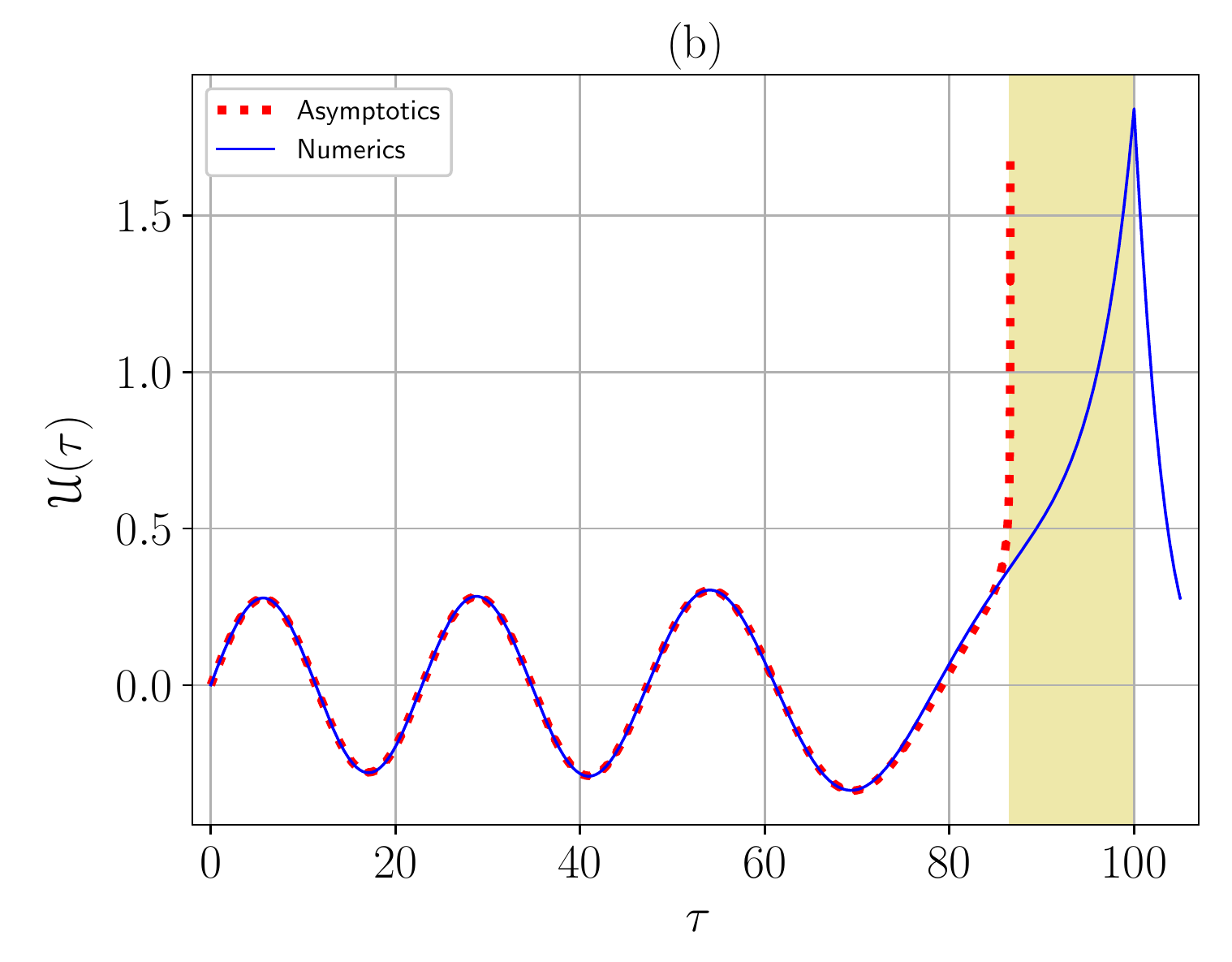}}
\caption{
Comparing the asymptotic solution in the form of 
Eqs.~\eqref{U_ne_0-1}--\eqref{P-analytic-free}
and the corresponding numerical solution for the accelerating oscillator
moving at the speed
$v=T$: $\epsilon=0.01$, $M=12$, $K=-1$. (a) The internal force, (b)
the displacement. 
The yellow span in each sub-plot corresponds to the second stage of the
motion, when the
solution is still sub-critical, but restriction \eqref{v-Kless0} is not
satisfied.}
\label{Mg0-Kl0.pdf}
\end{figure}

\afterpage{\clearpage}
In Fig.~\ref{Mg0-Kl0.pdf} we compare the results for the internal force  
$P(\tau)$ (a) and the
displacement $\mathscr U(\tau)$ (b) obtained for big enough mass
$M=12$.  The yellow span in each sub-plot corresponds to the time interval, where
the solution is still sub-critical, but restriction \eqref{v-Kless0} is not
satisfied. Thus, our asymptotic solution is defined only for the time values to
the left of the yellow span.  The left boundary of the span corresponds to the
instant when immediate value of the trapped mode frequency approaches  
zero (this corresponds to the string buckling).  The right boundary of the
span corresponds to the instant of overcoming the critical speed ($v=1$).

Again, the asymptotic solution approaches the numerical one very quickly. 
The numerical and asymptotic solutions diverge just at the left boundary of the
span. The asymptotics approaches infinity at the left boundary, whereas the
numerical solution demonstrates a growth within the span. At the right boundary of the
span the displacement becomes to be prescribed by the perturbations, radiated
in the past, during the sub-critical stage of the motion
\cite{Non-Stationary}. 
After that instant the displacement $\mathscr U(\tau)$ decreases.  

\subsubsection{Pure free oscillation in the case when the trapped mode does not exist initially}
\label{sec-not-exist}
Assume now that initially in the system with $v=v(0)$ condition 
\eqref{v-Kgt0} is not satisfied, and the trapped mode does not exist, i.e.,\ 
the constructed analytic solution is not valid. Thus, the first stage of the
motion does not exist.
Note that we have zero second
term describing the free oscillation in the corresponding system with constant
parameters (see Eq.~\eqref{itog}); therefore, we expect
that the free oscillation in the system with time-varying $v(T)$ is negligible.
We will discuss numerics for the practically more
important case $M>0,\ K>0$.
In Fig.~\ref{Mg0-Kg0-no.pdf} we present the numerical results obtained 
for a monotonically increasing $v(T)$. 
\begin{figure}[hp]
\centering{\includegraphics[width=0.85\textwidth]{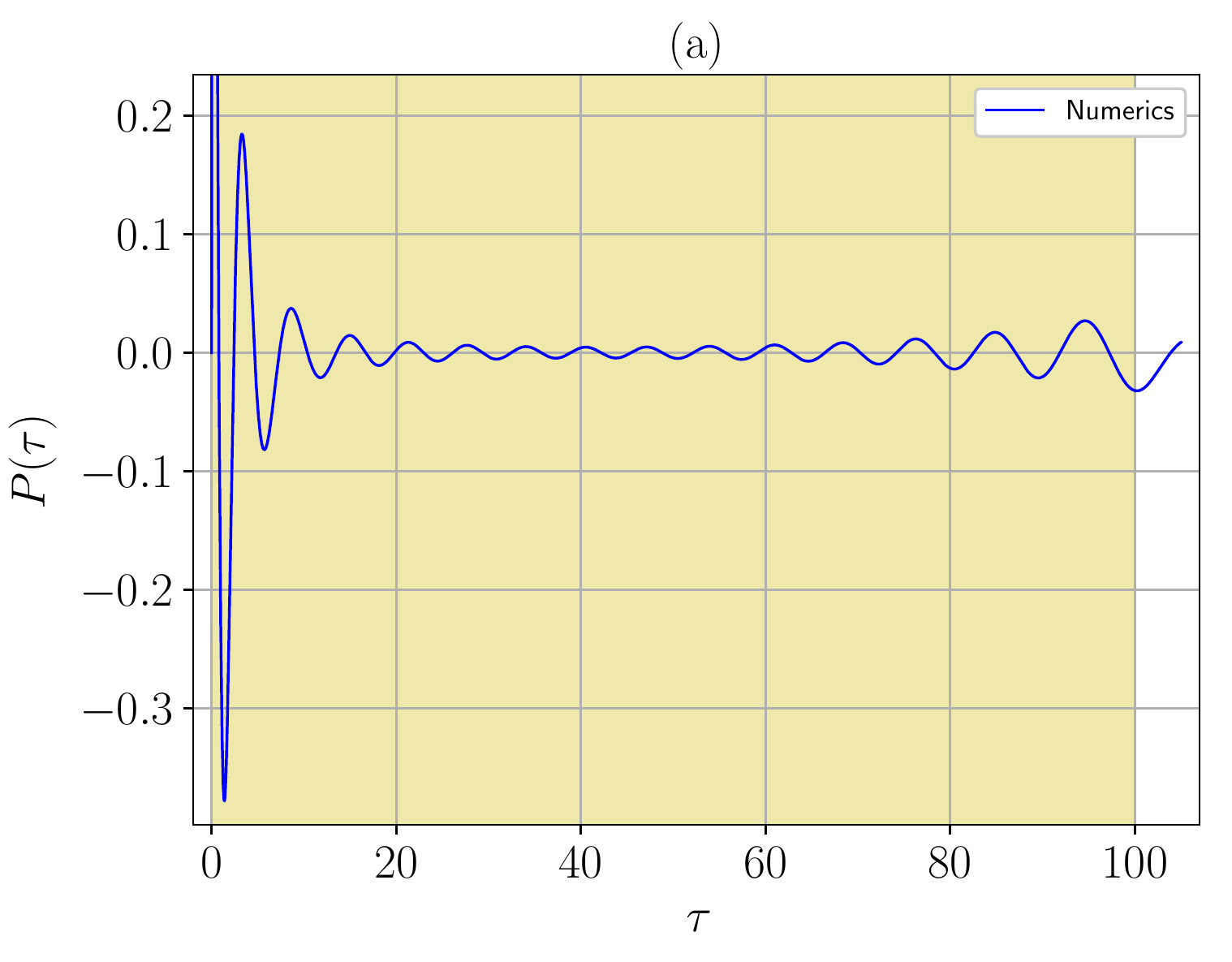}}
\centering{\includegraphics[width=0.85\textwidth]{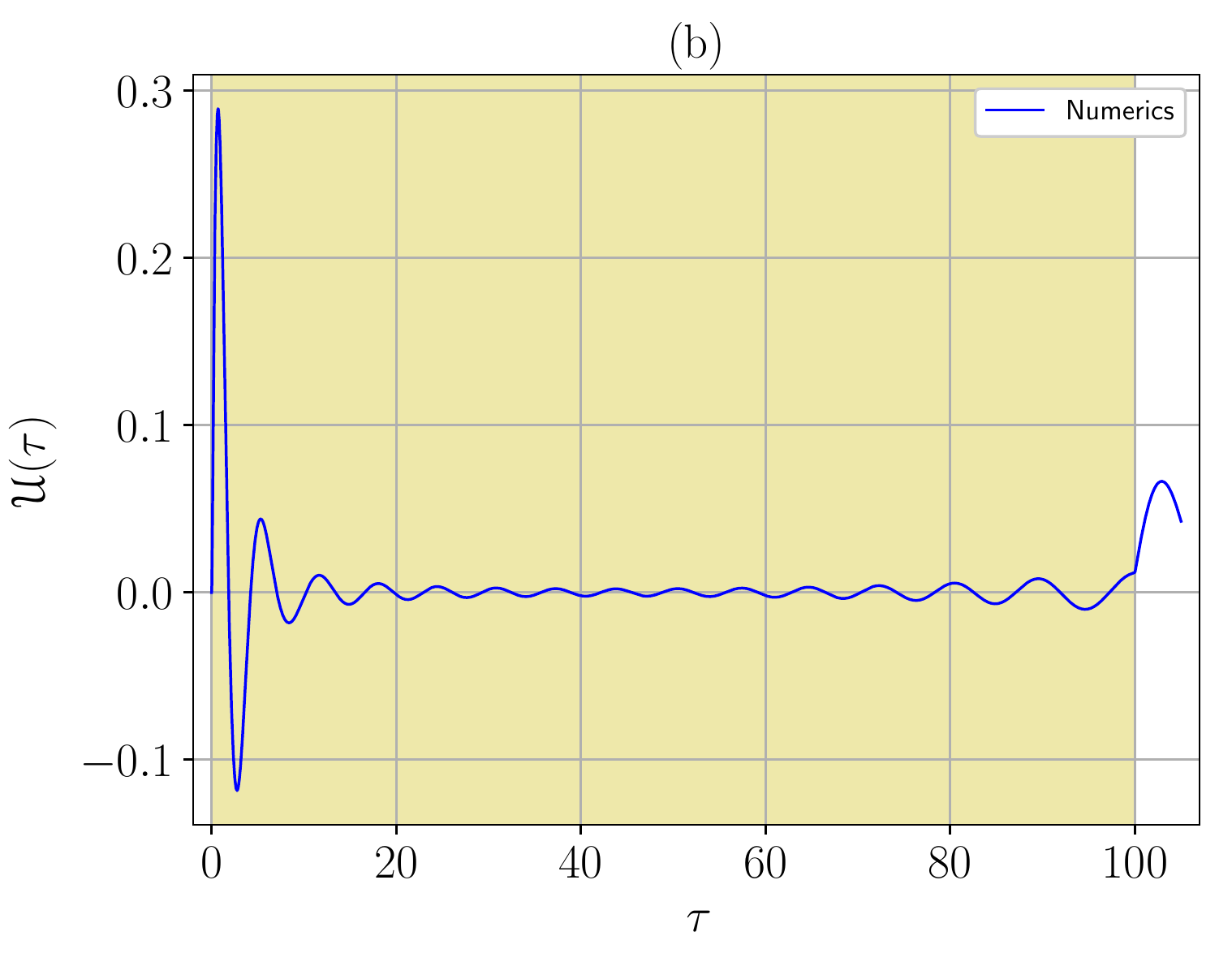}}
\caption{
Numerical solution for the accelerating oscillator moving at the speed
$v=T$: $\epsilon=0.01$, $M=1$, $K=3$. (a) The internal force, (b)
the displacement. The
yellow span in each sub-plot corresponds to the time interval, when the solution is still
sub-critical, but restriction \eqref{v-Kgt0} is not satisfied.}
\label{Mg0-Kg0-no.pdf}
\end{figure}
\afterpage{\clearpage}
One can see that free oscillation quickly vanishes as expected due to analysis
performed for the system with constant parameters. The amplitude of
oscillation is negligible comparing with the case, where the trapped mode
exists, and slightly increases before overcoming the critical speed $v=1$. 

\subsubsection{Free and forced oscillation}
\label{sec-forced}
{Now we take} 
\begin{equation}
 \hat p(\tau)=0,
 \qquad 
 N=1,
 \qquad 
 p^{(\Omega_1)}(T)\in\mathbb R,
\end{equation}
and, therefore, the external force $p(T,\tau)$ defined by Eq.~\eqref{p-def}
is as follows:
\begin{equation}
p(\tau,T)=2H(\tau)\,p^{(\Omega_1)}(T)\cos\left(
\int_0^\tau \Omega_1(T)\,\d T
\right).
\end{equation}
We put also
\begin{equation}
p^{(\Omega_1)}(T)=\alpha_0+\alpha_1 T,
\qquad
\Omega_1(T)=\gamma_0+\gamma_1 T.
\end{equation}
One has (see \eqref{ft-exp})
\begin{equation}
\mathscr{F}\{p(\tau,0)\}\big(\Omega_0(0)\big)=
p^{(\Omega_1)}(0)\,\frac{2\I\Omega_0(0)}{\Omega_0^2(0)-\Omega_1^2(0)}.
\label{Fp-cos}
\end{equation}
The last equation should be used to define the unknown constants 
\eqref{C0-def}, \eqref{D0-def}.

\begin{figure}[hp]%
\centering{\includegraphics[width=0.85\textwidth]{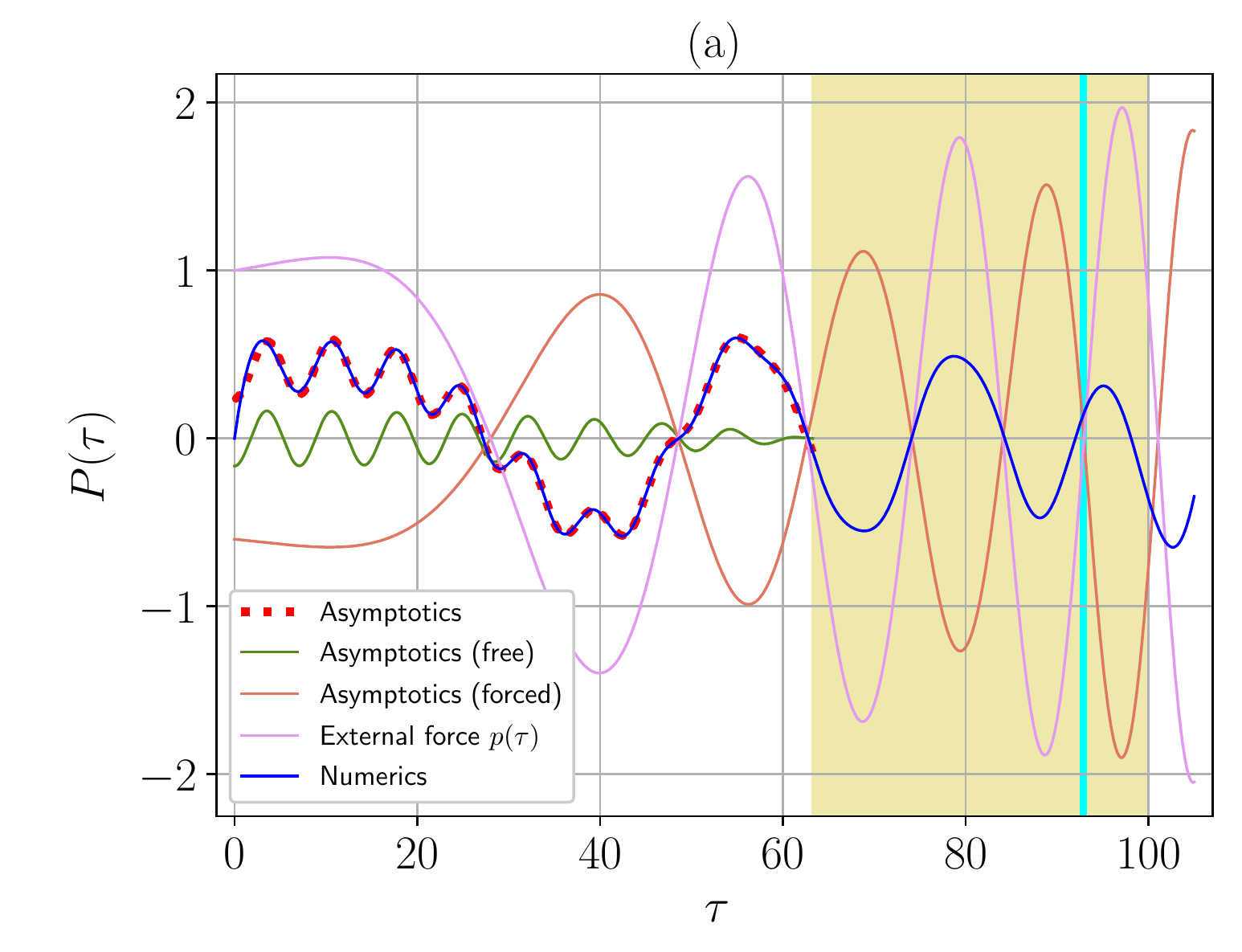}}
\centering{\includegraphics[width=0.85\textwidth]{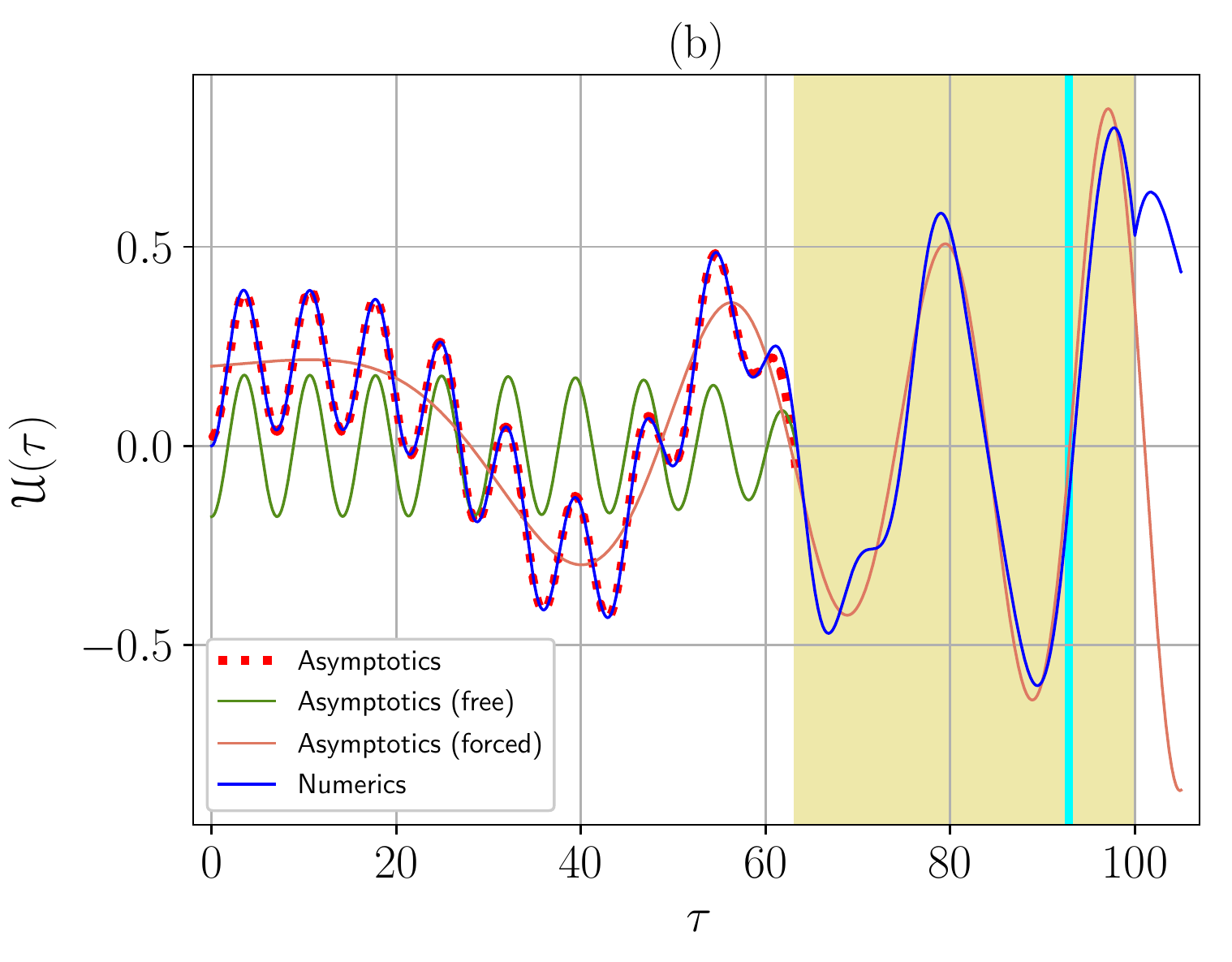}}
\caption{
Comparing the asymptotic solution in the form of 
Eqs.~\eqref{U_ne_0-1}--\eqref{P-analytic-free} represented as the superposition of the free
oscillation, the forced one, and the external force (for the plot (a))
with the corresponding numerical solution for the accelerating oscillator moving
at the speed $v=T$: $\epsilon=0.01$, $M=5$, $K=3$  in the case of
the low-frequency excitation. (a) The internal force, (b) the displacement.  The
yellow span in each sub-plot corresponds to the second stage of the motion, 
when the solution
is still sub-critical, but restriction \eqref{v-Kgt0} is not satisfied. The
cyan vertical line corresponds to the instant when the excitation frequency 
becomes equal to the cut-off frequency.}
\label{Mg0-Kg0-forced-low.pdf}
\end{figure}

\begin{figure}[hp]%
\centering{\includegraphics[width=0.85\textwidth]{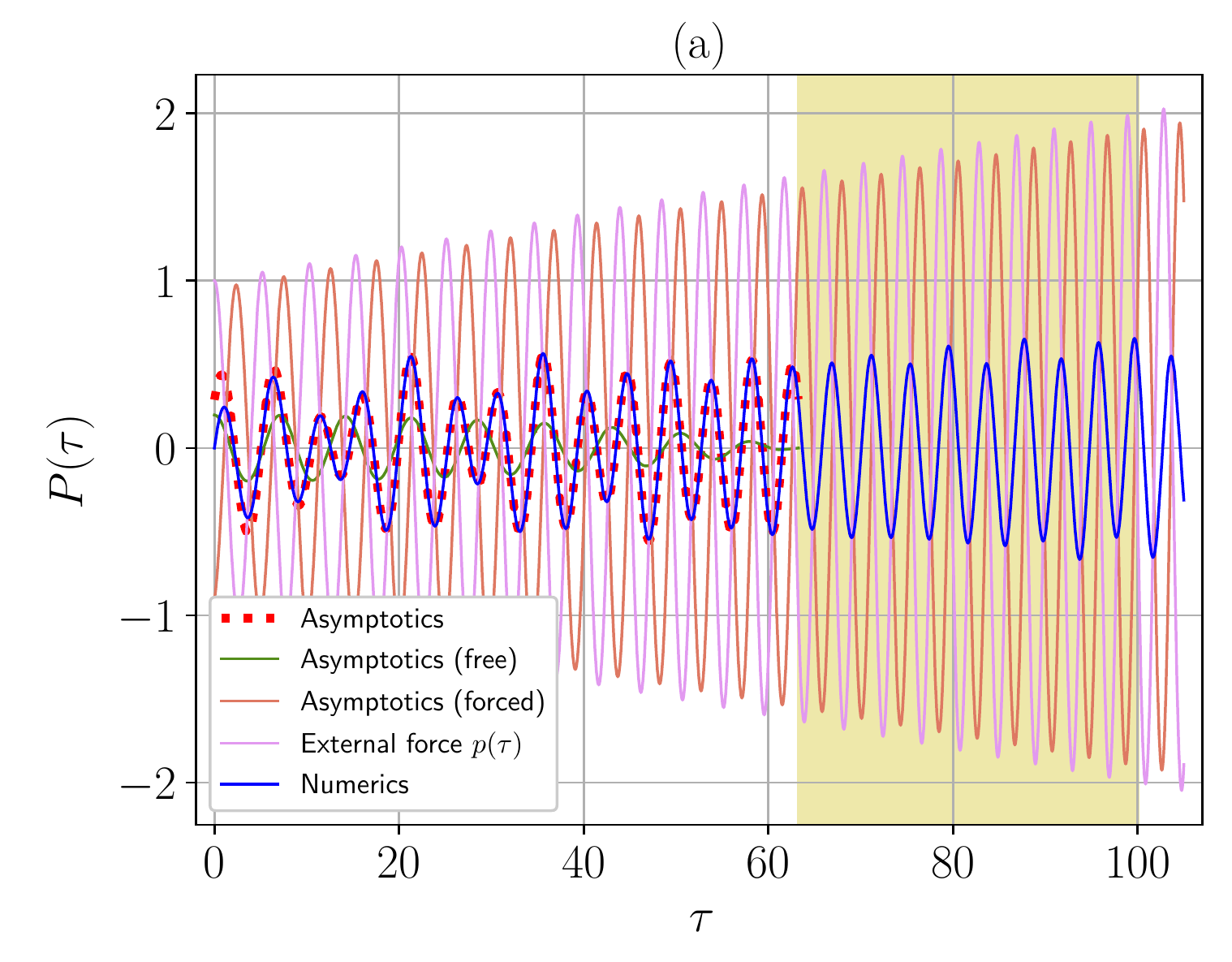}}
\centering{\includegraphics[width=0.85\textwidth]{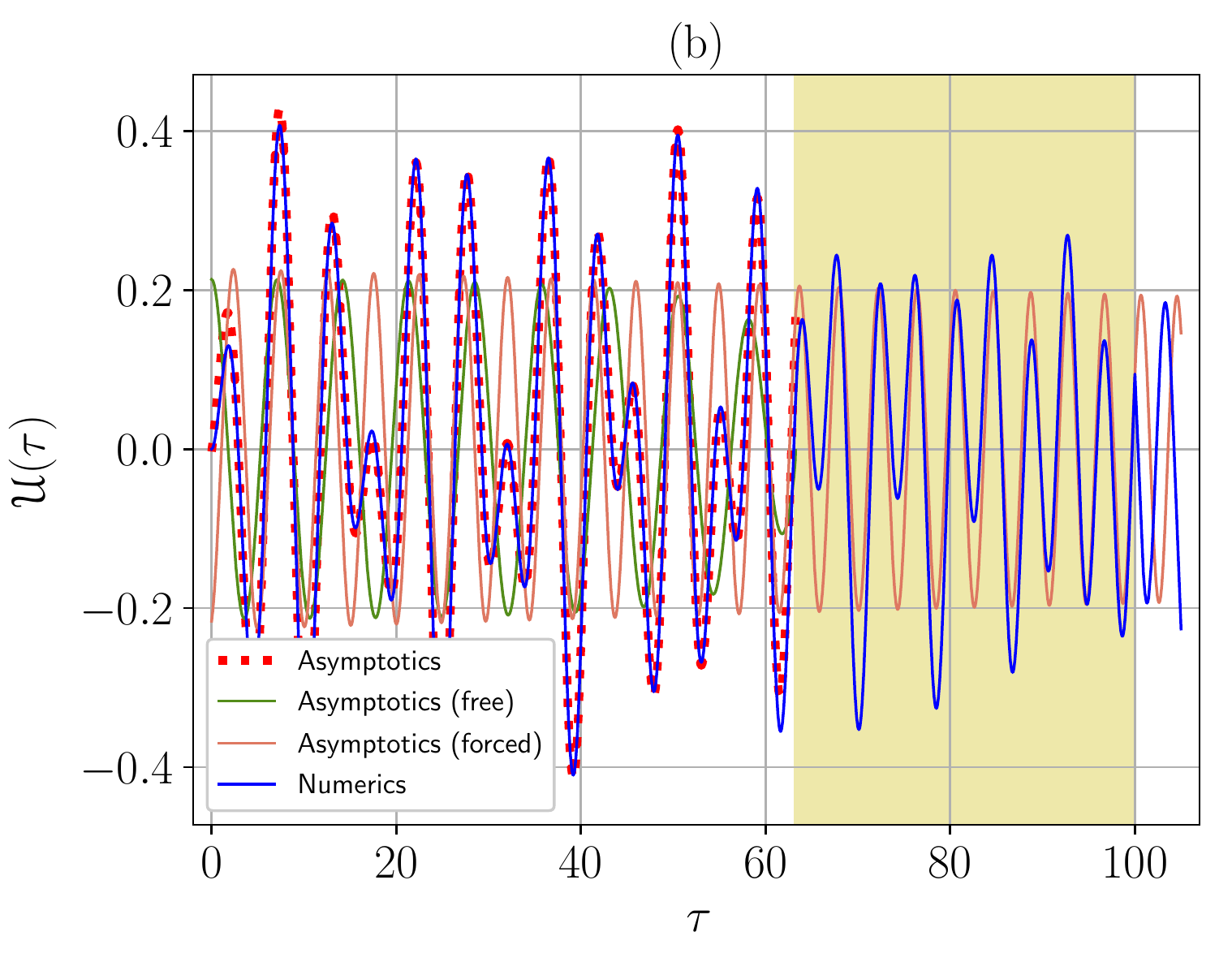}}
\caption{
Comparing the asymptotic solution in the form of 
Eqs.~\eqref{U_ne_0-1}--\eqref{P-analytic-free} represented as the superposition of the free
oscillation, the forced one, and the external force (for the plot (a))
with the corresponding numerical solution for the accelerating oscillator moving
at the speed $v=T$: $\epsilon=0.01$, $M=5$, $K=3$  in the case of
the high-frequency excitation. (a) The internal force, (b) the displacement.  The
yellow span in each sub-plot corresponds to the second stage of the motion,
when the solution
is still sub-critical, but restriction \eqref{v-Kgt0} is not satisfied. 
}
\label{Mg0-Kg0-forced-high.pdf}
\end{figure}

\begin{figure}[hptb]%
\centering{\includegraphics[width=0.8\textwidth]{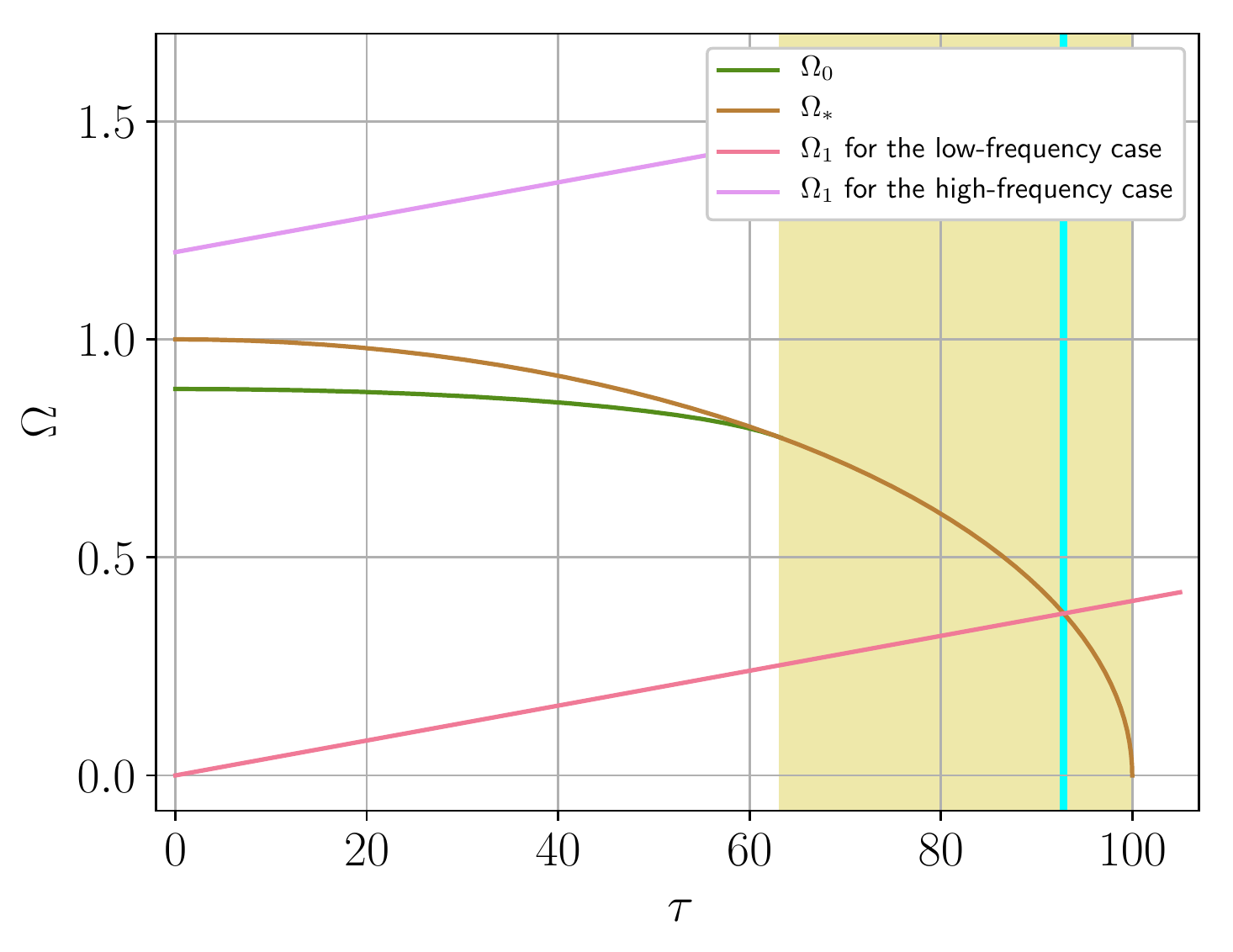}}
\caption{The trapped mode
frequency $\Omega_0$, the cut-off frequency $\Omega_\ast$, and the external
excitation frequencies $\Omega_1$ versus time $\tau=T/\epsilon$ for the
problems illustrated by Figs.~\ref{Mg0-Kg0-forced-low.pdf} 
\&
\ref{Mg0-Kg0-forced-high.pdf} (the low-frequency and the high frequency cases,
respectively).
The yellow span corresponds to the second stage of the motion, when the solution
is still sub-critical, but restriction \eqref{v-Kgt0} is not satisfied. 
The cyan vertical line
corresponds to the instant when the excitation frequency becomes equal to the
cut-off frequency (in the low-frequency case).}
\label{Mg0-Kg0-forced-Omega.pdf}
\end{figure}

We will compare the asymptotic and numerical results for the practically more
important case $M>0,\ K>0$,
and assume that initially in the system with $v=v(0)$ condition 
\eqref{v-Kgt0} is satisfied and the trapped mode exists.
Since we deal with a non-resonant excitation only
(for which \eqref{non-resonant0} is fulfilled for all $T$), there are two qualitatively
different cases, namely the low-frequency case $\Omega_1<\Omega_0$ (the forced
oscillation is localized near the mass-spring system) and the
high frequency case $\Omega_1>\Omega_0$ (the forced oscillation forms propagating
waves). In Figs.~\ref{Mg0-Kg0-forced-low.pdf} \&
\ref{Mg0-Kg0-forced-high.pdf} we compare the asymptotic and numerical results
obtained in the low frequency case, where we take
\begin{equation}
\alpha_0=0.5,\quad \alpha_1=0.5,\quad \gamma_0=0,\quad \gamma_1=0.4,
\end{equation}
and in the high frequency case, where 
\begin{equation}
\alpha_0=0.5,\quad \alpha_1=0.5,\quad \gamma_0=1.2,\quad \gamma_1=0.4,
\end{equation}
respectively.
In Fig.~\ref{Mg0-Kg0-forced-Omega.pdf} we present the plot for the trapped mode
frequency $\Omega_0$, the cut-off frequency $\Omega_\ast$, and the external
excitation frequencies $\Omega_1$ versus time $\tau=T/\epsilon$ for the
problems illustrated by Figs.~\ref{Mg0-Kg0-forced-low.pdf} \&
\ref{Mg0-Kg0-forced-high.pdf}. The yellow span in all these figures
corresponds to the time interval, where the solution
is still sub-critical, but restriction \eqref{v-Kgt0} is not satisfied. 
Thus, our asymptotic solution for the free oscillation is defined only for
the time values to the left of the yellow span.  One can see that the
asymptotic solution is in a very good agreement with the numerical one 
for such values of time.
The cyan vertical line
corresponds to the instant when the excitation frequency becomes equal the
cut-off frequency (in the low-frequency case).
According to Eq.~\eqref{P-analytic} the asymptotic solution for the internal force
$P(\tau)$ is the superposition of the free oscillation, the forced
oscillation, and the external force $p(\tau)$.  At the same time, the displacement
$\mathscr U(\tau)$ is the superposition of the free oscillation and the forced
oscillation. These individual components are also shown in the plots. One can
observe an excellent agreement between the analytic and numerical results. 

\subsection{Final remarks}
\label{sec-final-remarks}
\begin{remark}  
One can see in Fig.~\ref{Mg0-Kg0-forced-Omega.pdf} that restriction 
\eqref{non-resonant0}, which guaranties a non-resonant character of the
solution under consideration, is fulfilled for all admissible~$\tau$. 
If we break this requirement, e.g.,\ by taking of $\Omega_1(0)$ in the narrow
interval between the trapped mode frequency and the cut-off frequency, then
the constructed asymptotic solution becomes practically inapplicable.
\end{remark}

\begin{remark} 
The particular case, when $\Omega_1=0$ and the weight
$p=p^{(0)}_0=\mathrm{const}$, is very similar
to the case of the pure free oscillation considered in
Sect.~\ref{sec-free-gg}--\ref{sec-free-l}. According to 
Eqs.~\eqref{U-weight}, \eqref{U_ne_0-1}, \eqref{P-analytic}
the internal force and the
displacement are the superpositions of a free oscillation and a constant quantity. 
The unknown constants can be found by formulae 
\eqref{C0-def}, \eqref{D0-def}
and \eqref{Fp-cos} wherein $\Omega_1=0$. Due to 
Eqs~\eqref{D0-def}, \eqref{Fp-delta}, \eqref{Fp-cos}
the shift of the phase between
a free oscillation in the case under consideration and a free oscillation
considered in Sect.~\ref{sec-free-gg}--\ref{sec-free-l} is $\pi/2$.
\end{remark}

\begin{remark}
One can consider more complicated problems, where the external force is
a superposition of several harmonics. The possible practically important example
corresponds to the case, when both the weight and an external sinusoidal
excitation are applied to the mass-spring system. The corresponding
asymptotic solution will be in a good agreement with the numerical one.
\end{remark}

\begin{remark}
For the aim of simplicity, we have considered in
Sect.~\ref{sec-free-gg}--\ref{sec-forced} the test problems, wherein the quantities
$v(T)$, $\Omega_1(T)$, $p^{(\Omega_1)}(T)$, independently vary in a uniform way. One can
consider a more complicated problem, where these parameters vary in a
non-uniform, and even in a non-monotonous, way, keeping their values in the
corresponding admissible intervals. For enough small $\epsilon$, 
the corresponding asymptotic solutions
will be in a good agreement with numerical ones (an example of
non-monotonically changing parameters for another 
problem investigated by the same method can be found in \cite{arXiv:1805.07382}, see Fig.~3 there).
\end{remark}

\begin{remark}
For the aim of simplicity, we have considered in
Sect.~\ref{sec-free-gg}--\ref{sec-free-l} the test problems, wherein the
pulse loading is the function 
\eqref{rectangle} approximating $\delta(\tau)$.
One can
consider a more complicated problem, where 
this function even is not a finite, but an exponentially vanishing at infinity
function.
For enough small $\epsilon$, 
the corresponding asymptotic solution
will be in a good agreement with numerical one (an example of exponentially
vanishing pulse loading 
for another 
problem investigated by the same method can be found in \cite{arXiv:1805.07382}, 
see Fig.~4 there).
\end{remark}
\section{Conclusion} 
\label{sec-conclusion}
In the paper we have 
considered forced and localized free oscillation of a
string on the Winkler foundation subjected to a discrete mass-spring system
non-uniformly moving at a sub-critical speed.
We have used the analytic approach 
first time suggested in
\cite{gavrilov2002etm}. Now we have suggested
some mathematical trick, namely, transforming of the right-hand side of the first
approximation equation to the form of a total differential of the logarithm 
of a certain function, see Eq.~\eqref{tut-final}.
This allows us to significantly simplify calculations  
and obtain the analytic solution
of a quite complicated problem. 
We note that our analytic approach is not a pure
asymptotic one, since we guess which harmonics
we need to take into account in
the multi-frequency ansatz (see Sect.~\ref{sect-multi-frequency-ansatz}). 
Nevertheless, such an approach has a deep asymptotic motivation.
We successively apply two quite different asymptotic
methods: the method of stationary phase to the system with constant parameters and
the method of multiple scales to the system with time-varying parameters, and
then use a simple matching procedure to find unknown constants. It is very
difficult and perhaps impossible to prove that we obtain asymptotically
correct results in this way, thus, we need to verify the obtained results
numerically. This has been done in Sect.~\ref{sect-numerics}.

The specific feature of the solution describing oscillation of the string
subjected to a uniformly accelerated oscillator is that
there exist two qualitatively different stages of the 
motion. 
Indeed, assuming that the trapped mode initially exists, one can observe that
it disappears at a certain instant of time before overcoming the critical speed $v=1$. In the
case $K>0$, 
at this instant 
the trapped mode frequency $\Omega_0(T)$ approaches the cut-off frequency
$\Omega_\ast(T)$, which
is the boundary frequency for the continuous spectrum of natural frequencies. In
the case $K<0$, at this instant 
the trapped mode frequency approaches zero,
and we observe the dynamic instability and the buckling.

The main analytic result of the paper is
Eqs.~\eqref{U_ne_0-1}--\eqref{P-analytic-free}, which are valid in the absence
of resonance of any type (see Eqs.~\eqref{non-resonant0},
\eqref{non-resonant}). 
Herewith, the mass-spring system speed $v$,
the amplitudes $p^{(\Omega_i)}$ and the
frequencies $\Omega_i$ for the harmonics of the external force are assumed to
be independent slowly
time-varying functions. These formulae describe the first stage of the
sub-critical motion, where the non-vanishing free localized oscillation is
observed.  
We have demonstrated that the analytic solution is in a very good
agreement with the numerical one in the corresponding admissible interval of
the speed $v$. The free oscillation at the second stage is described only
numerically.
For the system with positive $K$, it has been shown
that the amplitude of the internal force $P$ 
can begin again to grow after the instant 
when the trapped mode disappears (but before  overcoming the critical speed),
see Fig.~\ref{Mg0-Kg0.pdf}.
This fact is a bit unexpected.
At the same time, the corresponding amplitude of the displacement $\mathscr U$ decreases
monotonically. 
For the system with negative $K$, numerical post-buckling solution grows until the
overcoming the critical speed $v=1$. After that the displacement $\mathscr U$
becomes to be prescribed by the
perturbations, radiated in the past \cite{Non-Stationary}, during the sub-critical stage of the
motion, and the displacement begins to decrease, see
Fig.~\ref{Mg0-Kl0.pdf}.
While considering the limiting case $K=0$, we have discovered  
an error in previous paper \cite{gavrilov2002etm} (see
Sect.~\ref{sect-comparison-old} for the corresponding discussion). Note that 
the erroneous solution obtained in \cite{gavrilov2002etm} has a behaviour, which
is very close to the correct solution. The dynamical behaviour of the system
considered in \cite{gavrilov2002etm} is much simpler, in particular the second
stage of a uniformly accelerated motion does not exist.

Finally, let us discuss how the results of the paper can be generalized. In
our opinion,  
to describe oscillation at the second stage 
in the case of the positive spring stiffness $K$
it
would be useful to construct a resonant solution describing overcoming the
cut-off frequency in spirit of \cite{Shishkina2020}, where a model resonant
solution for a system possessing a trapped mode is obtained. Nevertheless,
for the problem under consideration, the generalization of the results of 
\cite{Shishkina2020} does not seem to be
straight-forward. The solutions describing the passage through resonance,
where $\Omega_i\simeq\Omega_0$, apparently, also can be obtained in such a way.

The method used in the paper without essential changes can be applied to
another systems involving a string on the Winkler foundation and moving
discrete mass-spring systems, e.g.,\ ones considered in
\cite{Roy2018,Kruse1998}, under condition that a unique trapped mode initially
exists. 


\section*{Declaration of competing interest}
I.O.~Poroshin acknowledges the support of the Government of the Russian
Federation (state assignment 0784-2020-0027).
S.N.~Gavrilov acknowledges the support of the Russian Foundation for Basic
Research (grant 19-01-00633).


\section*{Acknowledgements}
The authors are grateful to Yu.A.~Mochalova for useful and stimulating discussions.

\appendix
\section{The system with constant parameters: basic assumptions}
\label{App-A}
In \ref{App-A}--\ref{neodnor}
we present some auxiliary  
results related to the governing equations in the form of 
Eqs.~\eqref{ur_str_3}, \eqref{ur_osc_2-dupe} in the case of constant
parameters.
In such a way we introduce several quantities and relations, which are
necessary in order to apply our asymptotic and numerical approaches.
The loading is assumed to be as follows:
\begin{gather}
v{}'_T\equiv0\quad\Longrightarrow\quad
v(T)\equiv v(T_0);
\label{vT0}
\\
\Omega_i{}'_T\equiv0,\quad i=\overline{1,N}
\quad\Longrightarrow\quad
\Omega_i{}(T)=
\Omega_i{}(T_0)
;\\
p^{(\Omega_i)}{}'_T\equiv0,\quad i=\overline{1,N}\quad\Longrightarrow\quad
p(\tau,T)\equiv p(\tau,T_0)
\end{gather}
for certain $T_0=\mathrm{const}$. The oscillator speed is assumed to be sub-critical, i.e.,
Eq.~\eqref{v-is-subcritical} is fulfilled.

\section{The dispersion relation for the Klein-Gordon equation in the moving
co-ordinates}
\label{aux-dispersion}
Consider properties of the linear differential operator in the left-hand side 
of Eq.~\eqref{ur_str_3} in the case \eqref{vT0}.
Assuming that $P(\tau)=0$ and
\begin{equation}
u(\xi,\tau) = \mathscr W \mathrm{e}^{-\I (\Omega \tau+\omega\xi)},
\end{equation}
we get the dispersion relation for the operator in the left-hand side of
Eq.~\eqref{ur_str_3} in the following form:
\begin{equation}
    \omega^2 - 2 B(\Omega) \omega + A^2 (\Omega) =0.
\label{disp-relation}
\end{equation}
Here $\omega$ is the wave-number,
\begin{gather}
    A^2(\Omega)\=\frac{1-\Omega^2}{\Omega_\ast^2},
\label{A-def}
\\    
    B(\Omega)\=\frac{\upsilon\Omega}{\Omega_\ast^2},
\label{B-def}
\end{gather}
where $\Omega_\ast$ is the cut-off frequency defined by Eq.~\eqref{cut-off}.
Put
\begin{equation}
S(\Omega)\=\sqrt{A^2(\Omega) - B^2(\Omega)} =
\frac{\sqrt{\Omega_\ast^2-\Omega^2}}{\Omega_\ast^2},
\label{S_Omega_v}
\end{equation}
where the principal branch of the square root is chosen.
Thus, from the dispersion relation~\eqref{disp-relation} one obtains the
expression for
the wavenumbers $\omega$:
\begin{equation}
        \omega=B(\Omega) \pm \I S(\Omega).
\label{wavenumber-expl}
\end{equation}
{One can see that according to dispersion relation~\eqref{disp-relation}
free waves with frequencies upper
than the cut-off frequency are sinusoidal propagating waves, whereas free waves 
with frequencies below than the cut-off frequency are growing inhomogeneous
waves, which cannot exist if we require boundedness}.

Note that quantity $B$ defined by Eq.~\eqref{B-def} satisfies equality:
\begin{equation}
\label{vB+O_0}
\upsilon B +\Omega= \frac{\Omega}{1-\upsilon^2}.
\end{equation}

\section{The Green function in the frequency domain for the system without
oscillator}
Consider now Eq.~\eqref{ur_str_3} and assume that
\begin{gather}
u(\xi,\tau) = W(\xi)\, \mathrm{e}^{-\I \Omega \tau},
\label{u-harmonic}
\\
P(\tau)=\mathrm e^{-\I\Omega\tau}.
\end{gather}

Let substitute expressions~\eqref{u-harmonic}
into Eq.~\eqref{ur_str_3}.  This yields
\begin{equation}
(1-v^2)\left(W''-2 \I B(\Omega)W'-A^2(\Omega)W\right)=
-
\delta(\xi),
\end{equation}
where $A^2(\Omega)$ and $B(\Omega)$ are defined by \eqref{A-def} and \eqref{B-def},
respectively.
The steady-state solution $W=G$ of the obtained equation is 
the Green function in the frequency domain for the system without
oscillator. This Green functions expresses the displacement of the string
subjected to a moving inertialess oscillating load. 
One can show that the Green function has the following form:
\begin{align}
&G(\xi, \Omega)=\frac{\exp\big(\I B\xi-S|\xi|\big)}{2\Omega_\ast^2\,S},&\quad
&|\Omega|<\Omega_\ast;
\label{Green-function-lower}
\\
&G(\xi,\Omega)=-\frac{\exp\big(\I B\xi-S\sign(\Omega)|\xi|\big)}
{2\Omega_\ast^2\sign(\Omega)\,S},&\quad
&|\Omega|>\Omega_\ast.
\label{Green-function-upper}
\end{align}
Expression 
\eqref{Green-function-lower} satisfies vanishing boundary conditions at
infinity, whereas Eq.~\eqref{Green-function-upper} satisfies the Sommerfeld
radiation conditions.

Note that the cut-off frequency $\Omega_\ast$ is a resonant frequency for the
system without oscillator: the amplitude of forced oscillation becomes
infinity as $\Omega\to\Omega_\ast$ due to the quantity $S$ defined by
Eq.~\eqref{S_Omega_v} in the denominator of the right-hand side of
Eq.~\eqref{Green-function-lower}. {The corresponding non-stationary solution
grows as $t\to\infty$}
\cite{slepyan1987energy,AyzenbergStepanenko2008,Abdukadirov_2019}. 

\section{The Green function in the frequency domain for the system with
oscillator}
\label{App-with}
Consider now Eqs.~\eqref{ur_str_3}, \eqref{ur_osc_2-dupe}, wherein 
\begin{gather}
p(\tau)=\mathrm e^{-\I\Omega\tau},
\end{gather}
assuming that Eq.~\eqref{u-harmonic} and
\begin{gather}
\mathscr{U}(\tau)=\mathscr W \mathrm{e}^{-\I \Omega \tau}
\label{harmonic-Uscr}
\end{gather}
are fulfilled.

Let substitute expressions~\eqref{u-harmonic}, \eqref{harmonic-Uscr}
into Eqs.~\eqref{ur_str_3}, \eqref{ur_osc_2-dupe}.  This yields
\begin{equation}
(1-v^2)\left(W''-2 \I B(\Omega)W'-A^2(\Omega)W\right)=
-\big(
(M \Omega^2-K)\mathscr{W}+1\big)\,
\delta(\xi),
\end{equation}
where $A^2(\Omega)$ and $B(\Omega)$ are defined by \eqref{A-def} and \eqref{B-def},
respectively.
The corresponding steady-state solution $W=\mathscr G$ is 
the Green function in the frequency domain for the system with the
oscillator. 
One can show that the Green function has the following form:
\begin{align}
&\mathscr G(\xi, \Omega)=\frac{\exp\big(\I
B\xi-S|\xi|\big)}{2\Omega_\ast^2\,S+K-M\Omega^2},&\quad
&|\Omega|<\Omega_\ast;
\label{Green-functionO-lower}
\\
&\mathscr G(\xi,\Omega)=-\frac{\exp\big(\I B\xi-S\sign(\Omega)|\xi|\big)}
{2\Omega_\ast^2\sign(\Omega)\,S+M\Omega^2-K},&\quad
&|\Omega|>\Omega_\ast.
\label{Green-functionO-upper}
\end{align}
Expression 
\eqref{Green-functionO-lower} satisfies vanishing boundary conditions at
infinity, whereas Eq.~\eqref{Green-functionO-upper} satisfies the Sommerfeld
radiation conditions.
Note that the trapped mode frequency $\Omega_0$ introduced in Sect.~\ref{spectral-joined}
is a root of the denominator 
in the right-hand side of Eq.~\eqref{Green-functionO-lower} according to
frequency equation~\eqref{ur_chastot}, and, therefore, this is a resonant
frequency.
Thus, the cut-off frequency
generally is not a resonant frequency for the system with oscillator.

\section{Spectral problem for a trapped mode}
\label{spectral-joined}

{Put $p = 0$ and consider the steady-state problem concerning the natural oscillations
of the system described by Eqs.~\eqref{ur_str_3}, \eqref{ur_osc_2-dupe},
assuming that 
Eqs.~\eqref{u-harmonic}, \eqref{harmonic-Uscr} are fulfilled.
We consider only sub-critical speeds, i.e.,
Eq.~\eqref{v-is-subcritical} is fulfilled.

A distributed system with discrete inclusions can possess a mixed spectrum of natural
frequencies
\cite{Ind-book-R2E,kaplunov2008example,Mishuris2020}.
In our case, clearly, there exists a continuous spectrum of frequencies, which lies
higher than the cut-off (or boundary) frequency: $|\Omega|\geq\Omega_\ast$.
Here $\Omega_\ast$ is given by Eq.~\eqref{cut-off}.
{The modes corresponding to the frequencies from the continuous spectrum are
harmonic waves.}
Trapped modes are modes with finite energy, therefore, we require
\begin{equation}
\int_{- \infty}^{+ \infty} W^2\, \mathrm{d}\xi< \infty,
\qquad 
\int_{- \infty}^{+ \infty} W'{}^2\, \mathrm{d}\xi< \infty.
\label{enery-restr}
\end{equation}
In the framework of the problem under consideration, 
they correspond to the frequencies from the
discrete part of the spectrum, which lies lower than the cut-off frequency:
$
0<|\Omega|< \Omega_\ast$.
We want to demonstrate that for the problem under consideration
under certain conditions the one and only one trapped mode with corresponding
frequency $\Omega_0>0$ can exist. 

Let substitute expressions~\eqref{u-harmonic}, \eqref{harmonic-Uscr}
into Eqs.~\eqref{ur_str_3}, \eqref{ur_osc_2-dupe}.  This yields
\begin{equation}
(1-v^2)\left(W''-2 \I B(\Omega)W'-A^2(\Omega)W\right)=
-(M \Omega^2-K)\mathscr{W}\,\delta(\xi),
\end{equation}
where $A^2(\Omega)$ and $B(\Omega)$ are defined by \eqref{A-def} and \eqref{B-def},
respectively.
The solution of the above equation can be  written as follows:
\begin{equation}
W(\xi)=(M\Omega^2-K)\mathscr{W}G(\xi, \Omega),
\label{W=W}
\end{equation}
where $G(\xi, \Omega)$ is the Green function 
\eqref{Green-function-lower}. This expression can be transformed to the
following equivalent form:
\begin{equation}
\label{resh2}
 W(\xi)=\frac{(M  \Omega ^2-{K}){\mathscr{W}}}{2\sqrt{1-\upsilon^2-\Omega^2}}
 \,\exp 
 \left(\frac{\I\Omega\upsilon\xi-\sqrt{1-\upsilon^2-\Omega^2}|\xi|}{1-\upsilon^2}\right).
\end{equation}
Calculating Eq.~\eqref{resh2}  at $\xi=0$ yields the
frequency equation for the trapped mode frequency $\Omega_0$ 
\begin{equation}
\label{ur_chastot}
	2\sqrt{\Omega_*^2-\Omega_0^2}= M \Omega_0 ^2-K,
\end{equation}
where 
\begin{equation}
\Omega_0^2>0.
\label{Omega2>0}
\end{equation}
Here Eq.~\eqref{cut-off} is taken into account. 

At first, consider the case $M>0$. Equation 
\eqref{ur_chastot}
with condition
\eqref{Omega2>0}
can be equivalently
rewritten as the system of the following bi-quadratic equation
\begin{equation}
\label{eq_sq_K<0}
	M^2\Omega_0^4 - 2(K M - 2)\Omega_0^2 +K^2 - 4\Omega_*^2=0
\end{equation}
together with inequalities 
\eqref{Omega2>0} and
\begin{gather}
\frac KM<\Omega_0^2
.
\label{MOV-osc-restr1-1}
\end{gather}
It follows from 
\eqref{ur_chastot},
\eqref{Omega2>0},
and
\eqref{MOV-osc-restr1-1} that 
\begin{equation}
0<\Omega_0^2<\Omega^2_\ast
\label{MOV-osc-restr1-2}
\end{equation}
and
\begin{gather}
0 \leq K<M\Omega^2_\ast  
\quad\text{if}\quad
K\geq0.
\label{restr-2}
\end{gather}

\begin{remark}  
Note that from inequality~\eqref{restr-2}
it follows that $ \Omega^2_\ast>K/M$, where $K/M$ is the squared partial frequency for the oscillator. 
\end{remark}

The discriminant for bi-quadratic equation 
\eqref{eq_sq_K<0} is
\begin{equation}
D= \frac{16(\Omega_*^2 M^2 -K M + 1)}{M^4}.
\label{discri}
\end{equation}
The discriminant is positive if and only if
\begin{equation}
K<\frac{1}{M}+M\Omega^2_\ast.
\end{equation}
This inequality is clearly true for $K<0$, and it is true
due to the second inequality in \eqref{restr-2} for $K>0$. Thus, the
frequency equation  
\eqref{ur_chastot}
is equivalent to the system of the following equation for the squared
frequency:
\begin{equation}
\label{3}
\Omega^2_{0(\pm)} =\frac{K M-2 \pm 2\sqrt{\Omega_*^2 M^2-K M+1}}{M^2}.
\end{equation}
and inequalities  \eqref{Omega2>0}, \eqref{MOV-osc-restr1-1}.

\begin{proposition}	
Provided that $M>0$, $K\geq0$
the root
$\Omega_0^2{}_{(+)}$ satisfies  inequalities~
\eqref{Omega2>0}, \eqref{MOV-osc-restr1-1}
if and only if $K$ satisfies inequality
\eqref{restr-2},
and the root
$\Omega_0^2{}_{(-)}$ does not satisfy inequality
\eqref{MOV-osc-restr1-1}. Thus, the trapped mode
exists and unique if 
\eqref{restr-2} is true, and does not exist otherwise.
\label{prop-1}
\end{proposition}
\begin{proof}	

If 
\eqref{restr-2} is not true, then
\eqref{MOV-osc-restr1-1} is not true and frequency equation 
\eqref{ur_chastot} does not have any positive roots $\Omega_0^2$.

On the other hand, 
let us prove that 
if \eqref{restr-2}
is true, then Eqs.~\eqref{Omega2>0}, \eqref{MOV-osc-restr1-1} 
wherein $\Omega_0^2=\Omega_0^2{}_{(+)}$
is true. 
Since \eqref{Omega2>0} follows from
\eqref{MOV-osc-restr1-1}, we need to check \eqref{MOV-osc-restr1-1} only.
We substitute $\Omega_0^2{}_{(+)}$ given by 
\eqref{3} into Eq.~\eqref{MOV-osc-restr1-1} and get
\begin{multline}
		\frac{K}{M} < \frac{K M-2 + 2\sqrt{\Omega_*^2 M^2-K M+1}}{M^2}
\\
\quad\Longleftrightarrow\quad
K M < K M-2 + 2\sqrt{\Omega_*^2 M^2-K M+1} 
\\
\quad\Longleftrightarrow\quad
1 <  \sqrt{\Omega_*^2 M^2-K M+1}.
\label{ineq-K>0-1}
\end{multline}
Since discriminant 
\eqref{discri}
is positive ($D>0$) provided that \eqref{restr-2}
is true, the last inequality is equivalent to 
Eq.~\eqref{restr-2}.

%
For $\Omega^2_{0(-)}$ one gets
\begin{equation}
K M<K M-2 - 2\sqrt{\Omega_*^2 M^2-K M+1},
\end{equation}
and this inequality is obviously not true.

\end{proof}	

\begin{proposition}	
Provided that $M>0$, $K<0$
the root
$\Omega_0^2{}_{(+)}$ satisfies 
inequalities  \eqref{Omega2>0}, \eqref{MOV-osc-restr1-1}
if and only if $K$ satisfies inequality
\begin{equation}
 -2|\Omega_\ast|<K<0,
\label{MOV-domain}
\end{equation}
and the root
$\Omega_0^2{}_{(-)}$ does not satisfy inequality
\eqref{MOV-osc-restr1-1}. Thus, the trapped mode
exists and unique if 
\eqref{MOV-domain}
is true, and does not exist otherwise.

\label{prop-2}
\end{proposition}

\begin{proof}	
Since \eqref{MOV-osc-restr1-1}
follows from \eqref{Omega2>0} we need to check \eqref{Omega2>0} only.
Let us prove that 
Eq.~\eqref{Omega2>0}
(wherein $\Omega_0^2=\Omega_0^2{}_{(+)}$)
is true if and only if 
\eqref{MOV-domain}
is true: 
%
\begin{multline}
0<K M-2 + 2\sqrt{\Omega_*^2M^2-K M+1}
\\
\quad\Longleftrightarrow\quad
(2-K M)^2<4(\Omega_*^2M^2-K M+1).
\\
\quad\Longleftrightarrow\quad
K^2<4\Omega_\ast^2.
\end{multline}
Since $K<0$, the last inequality is equivalent to \eqref{MOV-domain}.

The root $\Omega^2_{0(-)}$ clearly does not satisfy Eq.~\eqref{Omega2>0}.
\end{proof}


Consider  the special case $M=0$. The frequency equation~\eqref{ur_chastot} 
can be equivalently rewritten as follows
\begin{gather}
\Omega_0^2=\Omega_{*}^2-\frac{K^2}{4},
\label{freq-M=0}
\\
K<0.
\end{gather}
\begin{proposition}	
Provided that $M=0$, $K<0$
the root
\eqref{freq-M=0}
satisfies  inequality
\eqref{Omega2>0}
if and only if $K$ satisfies inequality \eqref{MOV-domain}.
Thus, the trapped mode
exists and unique if 
\eqref{MOV-domain}
is true, and does not exist otherwise.
\end{proposition}

\begin{proof}   
Inequality $\Omega_{*}^2-\frac{K^2}{4}>0$ is clearly equivalent to
Eq.~\eqref{MOV-domain}.
\end{proof}



Finally, the conditions for the existence of the trapped mode are
given by inequalities~
\eqref{restr-2} and
\eqref{MOV-domain} for $K\geq0$ and $K<0$, respectively.
These conditions can be written in the following equivalent forms in the terms
of the speed $v$:
\begin{gather}
\upsilon^2 < 1-\frac{K}{M}\quad \mathrm{if}\quad   M>0\ \mathrm{and}\ K \ge 0 ;
\label{v-Kgt0}
\\
\upsilon^2 < 1- \frac{K^2}{4}\quad \mathrm{if}\quad M\geq0\ \mathrm{and}\ K<0.
\label{v-Kless0}
\end{gather}

Consider the limiting case $v^2 =1-K/M-\lambda$, where $\lambda>0$ is a formal
small parameter. One can see that due to Eq.~\eqref{cut-off} the squared cut-off
frequency $\Omega^2_\ast= K/M+\lambda$. The asymptotics for the squared trapped mode frequency 
is
\begin{equation}
\Omega^2_0=\frac{K}{M}+\lambda-\frac{1}{4}\lambda^2 M^2+o(\lambda^2).
\end{equation}
Thus, a zone between the trapped mode frequency and the cut-off frequency is quite
narrow (see~Fig.~\ref{Mg0-Kg0-forced-Omega.pdf}).

Note that for the special case $K=0$ the critical value of the moving load speed
is $v=1$ (the speed of sound).
One can see that for $K \neq
0$ the critical value of $v$ is less than the speed of sound. {The special case
$K=0$ was considered by S.N.~Gavrilov and D.A.~Indeitsev in  \cite{gavrilov2002etm}. The
special case $v=0$, $M\ne0$ was considered in  \cite{gavrilov-da70,Glushkov2011a}.} 
The special case $v=0$, $M=0$ was considered in  \cite{arXiv:1805.07382}.} 

%
%

\section{Non-stationary free and forced oscillation}
\label{neodnor}
Consider the case when $T_0=0$, $p(\tau,T)\equiv p(\tau,0)\neq0$. Applying the
Fourier transform with respect to time $\tau$,%
\footnote{We understand the Fourier transform here as the
Fourier transform for generalized functions (or distributions)
\cite{Lighthill1964,Vladimirov1971}.} we get 
\begin{equation}
\mathscr{F} \{ u \}(\xi, \Omega)
=
\mathscr{F}\{p(\tau,0)\}\,
\mathscr G(\xi, \Omega),
\end{equation}
where symbol $ \mathscr{F} \{\cdot\}$ denotes the Fourier transform of the
corresponding quantity.
Applying the inverse transform yields
\begin{equation}
\label{1.1}
\mathscr U= \frac{1}{2\pi} \int_{-\infty}^{+\infty}
\frac{\mathscr{F} \{ p(\tau,0) \}\, \mathrm{e}^{-\I \Omega \tau} \, \d \Omega}
{2\sqrt{1-\upsilon^2- \Omega^2}+K -
M\Omega^2},
\end{equation}
where 
\begin{equation}
\mathscr{F}\{p(\tau,0)\}=\mathscr{F}\{\hat p\}
+\sum_{i=1}^N
\left(
p^{(\Omega_i)}(0)\,\frac{\I}{\Omega-\Omega_i+\I0}
-
\bar p^{(\Omega_i)}(0)\,\frac{\I}{-\Omega-\Omega_i-\I0}
\right)
\label{ft-exp}
\end{equation}
due to \eqref{p-def} \cite{Brychkov1977}. Here and in what follows accent $(\bar\cdot)$
denotes the complex conjugation.

Now we want to apply the method of stationary phase to integral  
\eqref{1.1}. We are interested to obtain the non-vanishing as $\tau\to\infty$
terms. These terms are defined by contributions from infinitesimal
neighbourhoods of poles just below the real axis \cite{fedoruk1977}. 
Note that $\mathscr{F}\{\hat p\}$ does not have any poles since $\hat p$
{is a finite function}.
{There are
poles of $\mathscr F\{p(\tau,0)\}$ at $\Omega=\pm\Omega_i-\I0$ and, due to
\eqref{ur_chastot}, there are
poles of the denominator of the right-hand side of Eq.~\eqref{1.1} at
$\Omega=\pm\Omega_0$.} The latter couple of poles must be
addressed in a special way to find their positions with respect to the real
axis. To do this we use the limiting absorption principle in the same way as it
was done {in \cite{Non-Stationary}}. We have to add the dissipative viscous term into
governing equation \eqref{ur_str_2-dim}, repeat all the calculations and find
the roots 
of the denominator for right-hand side of Eq.~\eqref{1.1} modified in such a way. Then we
consider a limiting case of zero friction to define the positions of the poles 
with respect to the real axis. One can show that the poles are shifted into
the lower half-plane of the complex plane: $\Omega=\pm\Omega_0-\I0$.
We apply the residue theorem,
Jordan's lemma, and the method of stationary phase to asymptotic evaluation of
the integral in 
the right-hand side of \eqref{1.1}. This results in
\cite{fedoruk1977,Non-Stationary}
\begin{equation}
\mathscr U = 
- \I \sum_{i=0}^N\sum_{\hat{\Omega}=\pm \Omega_i} 
\operatorname{Res} 
\left( \frac{\mathscr{F} \{ p(\tau,0) \}}{2\sqrt{1-\upsilon^2- \Omega^2}+K - M\Omega^2}, \hat{\Omega} \right)
\mathrm{e}^{-\I\hat{\Omega} \tau} + o(1), \quad \tau \to \infty.
\end{equation}
One has 
\begin{gather}
\text{Res} \left( \frac{1}{2\sqrt{1-\upsilon^2- \Omega^2}+K - M\Omega^2}, 
\pm\Omega_0 \right)
 = \mp \frac{\sqrt{1-\upsilon^2-\Omega_0^2}}{2\Omega_0
\left(1+M\sqrt{1-\upsilon^2-\Omega_0^2} \right)},
\\
\text{Res}\big(\mathscr{F}\{p(\tau,0)\},\Omega_i\big)=
\I p^{(\Omega_i)}(0),
\\
\text{Res}\big(\mathscr{F}\{p(\tau,0)\},-\Omega_i\big)=
\I \bar p^{(\Omega_i)}(0).
\end{gather}
Finally, we obtain the following asymptotic formula for the displacements:
\begin{multline}
\label{itog}
\mathscr U = 
2\sum_{i=1}^N |p^{(\Omega_i)}(0)\,\mathscr G(0,\Omega_i)|
\cos\Big(\Omega_i\tau-
\arg\big(p^{(\Omega_i)}(0)\,\mathscr G(0,\Omega_i)\big)\Big)
\\
+\frac{\sqrt{1-\upsilon^2-\Omega_0^2}
\,\big|\mathscr{F} \{ p(\tau,0) \}(\Omega_0)\big|}
{\Omega_0 \left(1+M\sqrt{1-\upsilon^2-\Omega_0^2} \right)} 
\sin \Big( \Omega_0 \tau - \arg \big(\mathscr{F} \{ p(\tau,0) \}(\Omega_0)\big)\Big)+o(1),
\quad \tau \to \infty.
\end{multline}

{Hence, for the large times, the
non-stationary response of the system under consideration is the
superposition of the undamped
modes of the forced oscillation
with frequencies $\Omega_i, i=\overline{1,N}$} (the first term in \eqref{itog})
and the undamped mode with the trapped mode frequency $\Omega_0$ 
(the second term in \eqref{itog}).

\section{The fundamental solution for the Klein-Gordon PDE}
\label{app-f-s}
Consider Eq.~\eqref{ur_str_2-dim} in the case $P(t)=\delta(t)$, $\ell(t)=0$
with initial conditions \eqref{initial-cond}. The corresponding solution
$u=\Phi$ is 
the Green function in time domain for the Klein-Gordon equation (the
fundamental solution). It has the form (see, e.g.,~\cite{polyanin2002handbook}):
\begin{equation}
\Phi(x,t)=\frac 12 H\big(t-|x|\big)J_0\big(\sqrt{t^2-x^2}\big).
\label{fund-sol}
\end{equation}
Here $J_0(\cdot)$ is the Bessel function of the first kind of zero order.

\section{The fundamental solution for the ODE describing a linear oscillator}
\label{app-f-s1}
Consider a free linear oscillator described by Eq.~\eqref{ur_osc_2}
wherein $P(t)=0$, $M>0$, which is subjected to the force $p(t)=\delta(t)$. The initial
conditions are formulated in the following form: $\mathscr U\big|_{t<0}=0$.
The corresponding solution $\mathscr U(t)=\Psi(t)$ is the fundamental solution
for the ODE describing a linear oscillator \cite{Vladimirov1971}:
\begin{equation}
\Psi(t)=\left[
\begin{aligned} 
&\frac{H(t)\sin \left(   \sqrt{\frac{K}{M}}\,t\right)}{\sqrt{K M}} , &\quad&K>0,
\\
&\frac{H(t)\sinh \left(   \sqrt{\frac{-K}{M}}\,t\right)}{\sqrt{-K M}} , &\quad&K<0,
\\
&\frac{H(t)\,t}{{M}} 
, &\quad&K=0.
\end{aligned}
\right.
\label{fund-sol-osc}
\end{equation}


\bibliographystyle{elsarticle-num}
\biboptions{compress}
\bibliography{string-spring,mode-trans,math,serge-gost,in-library,mode,discrete,journals.rus,moving_load,string,formulation,metamat,pantograph}

\end{document}